\documentclass[12pt,preprint]{aastex}
\usepackage{graphics}
\input epsf
\begin{document}

\title{Galaxy Morphologies in the Hubble Ultra Deep
Field: Dominance of Linear Structures at the Detection Limit}

\author{Debra Meloy Elmegreen \affil{Vassar College,
Dept. of Physics \& Astronomy, Box 745, Poughkeepsie, NY 12604;
e-mail: elmegreen@vassar.edu} }

\author{Bruce G. Elmegreen \affil{IBM Research Division, T.J. Watson
Research Center, P.O. Box 218, Yorktown Heights, NY 10598, USA; e-mail:
bge@watson.ibm.com} }

\author{Douglas S. Rubin \affil{Wesleyan University,
Dept. of Astronomy, Middletown, CT;
e-mail:rubin@wesleyan.edu}}

\author{Meredith A. Schaffer \affil{Vassar College,
Dept. of Physics \& Astronomy, Box 745, Poughkeepsie, NY 12604;
e-mail:meschaffer@vassar.edu}}

\begin{abstract}
Galaxies in the Hubble Ultra Deep Field (UDF) larger than 10
pixels (0.3 arcsec) have been classified according to morphology
and their photometric properties are presented. There are 269
spirals, 100 ellipticals, 114 chains, 126 double-clump, 97
tadpole, and 178 clump-cluster galaxies. We also catalogued 30
B-band and 13 V-band drop-outs and calculated their star formation
rates. Chains, doubles, and tadpoles dominate the other types at
faint magnitudes. The fraction of obvious bars among spirals is
$\sim10$\%, a factor of 2-3 lower than in other deep surveys.  The
distribution function of axial ratios for elliptical galaxies is
similar to that seen locally, suggesting that ellipticals relaxed
quickly to a standardized shape.  The distribution of axial ratios
for spiral galaxies is significantly different than locally,
having a clear peak at $\sim0.55$ instead of a nearly flat
distribution.  The fall-off at small axial ratio occurs at a
higher value than locally, indicating thicker disks by a factor of
$\sim2$.  The fall-off at high axial ratio could be from intrinsic
triaxial shapes or selection effects.  Inclined disks should be
more highly sampled than face-on disks near the surface brightness
limit of a survey. Simple models and data distributions
demonstrate these effects. The decreased numbers of obvious spiral
galaxies at high redshifts could be partly the result of surface
brightness selection.
\end{abstract}


\section{Introduction}

Galaxies with unusual morphologies appear at high redshift in the
Hubble Deep Fields North (Williams et al. 1996) and South
(Volonteri, Saracco, \& Chincarini 2000), the Hawaiian Deep Field
(Cowie, Hu, \& Songaila 1995), the Tadpole Galaxy field (Tran et
al. 2003), the GOODS field (Giavalisco et al. 2004) and the Hubble
Ultra Deep Field (HDF; Beckwith et al. 2005). Some are
recognizable as elliptical and spiral galaxies, but many at
$z\geq1.5$ have more clumpy and irregular structures (e.g.,
Abraham et al. 1996a,b; Conselice 2004).  Chain galaxies, for
example, are nearly straight alignments of a half-dozen clumps
(Cowie, Hu, \& Songaila 1995; van den Bergh et al. 1996;
Elmegreen, Elmegreen \& Sheets 2004, hereafter Paper I). Tadpole
galaxies are curved thin structures with a big clump near one end
(van den Bergh et al. 1996; Paper I; Straughn et al. 2004).
Luminous diffuse objects are oval distributions of clumpy emission
resembling disks (Reshetnikov, Dettmar, \& Combes 2003; Conselice
et al. 2004); a subcategory of these, having no bulges or
exponential disk light profiles, has been called clump-clusters
(Elmegreen, Elmegreen, \& Hirst 2004a, hereafter Paper II).
Double-clump galaxies have two big clumps. Most of these types are
not present in the modern Universe although some probably evolve
into normal Hubble types. Characteristic of all the irregular
types are enormous clumps of star formation, $\sim100$ times more
massive the the largest star complexes in today's spiral galaxies,
as well as thick disks and extreme asymmetries that are most
likely the result of interactions and merging.

Here we present morphological data on 884 galaxies in the Hubble
Ultra Deep Field (UDF). Magnitudes, colors, and axial ratios of
the galaxies are studied.

\section{Data}

UDF images were obtained with the Hubble Advanced Camera for
Surveys (ACS) by Beckwith et al. (2005) and are available on the
Space Telescope Science Institute (STScI) archive. The images are
10500 x 10500 pixels with a scale of 0.03 arcsec per px (315
arcsec x 315 arcsec) in 4 filters: F435W (B band, hereafter
B$_{435}$; 134880 s exposure), F606W (V band, V$_{606}$; 135320
s), F775W (i band, i$_{775}$; 347110 s), and F850LP (z band,
z$_{850}$; 346620 s).

\begin{figure}\epsscale{1.}
\plottwo{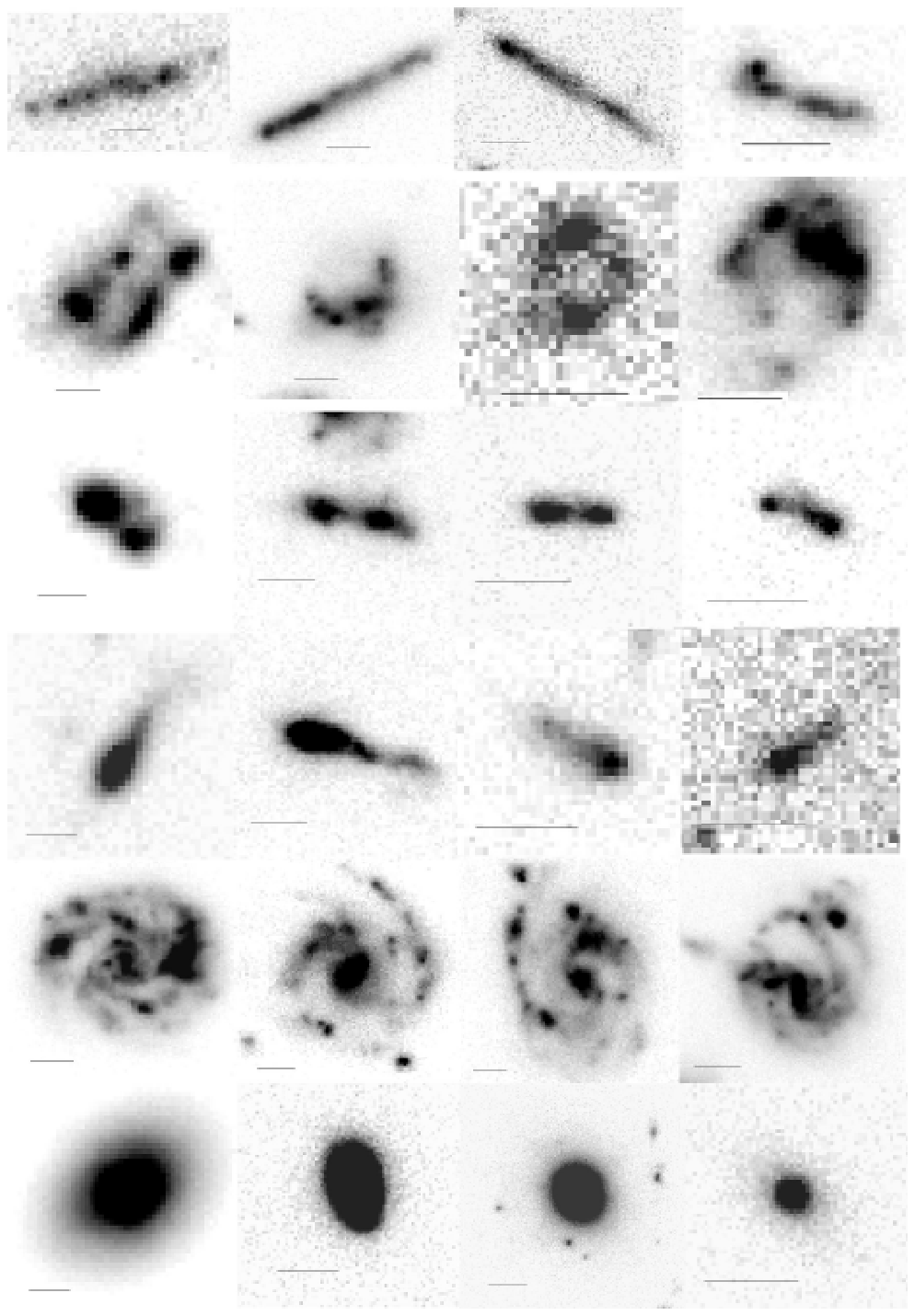}{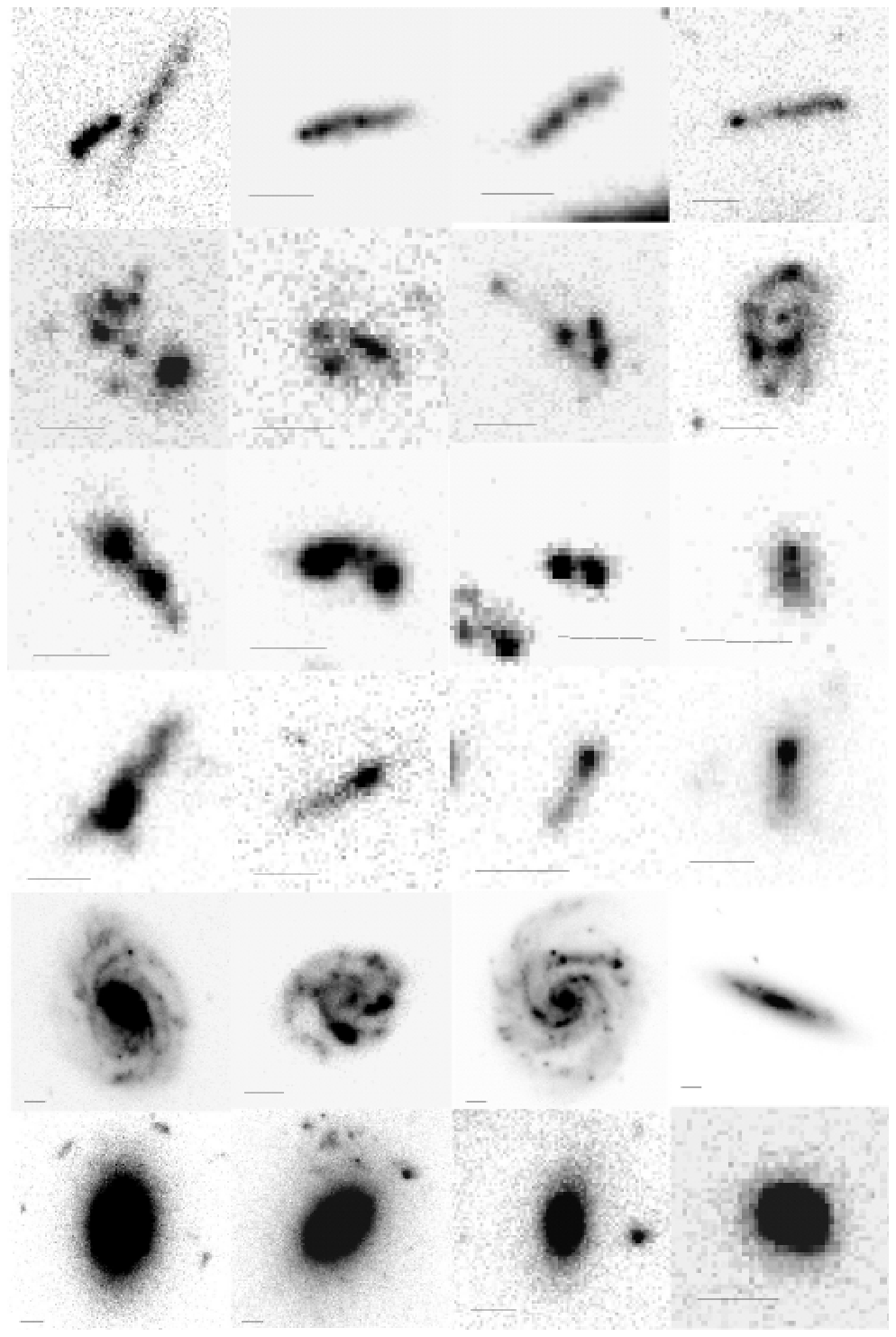} \caption{A selection of 8 typical
galaxies for each morphological type: 4 in Fig a and 4 in Fig b.
From top to bottom: chain, clump-cluster, double, tadpole, spiral,
and elliptical. Images are at i$_{775}$-band, with a line
representing 0.5 arcsec. UDF or our own identification numbers
from left to right in Fig a are as follows: chains: 6478, 7269,
6922, 3214; clump clusters: CC12, 1375, 2291, 5190; doubles: 637,
4072, 5098, 5251; tadpoles: 3058, 8614, 5358, 6891; spirals: 3372,
3180, 4438, 8275; ellipticals: 2107, 4389, 2322, 4913. In Fig b,
the identifications are: chains: 169 and 170 (2 separate
galaxies), 1428, 401, 3458+3418; clump clusters: 6486, 4807, 7230,
9159; doubles: 2461, 2558, 4097, 3967; tadpoles: 9543, 5115, 3147,
9348; spirals: 2607, 5805, 7556, 5670; ellipticals: 8, 4527, 4320,
5959. Fig. b has an example of an edge-on
spiral.}\label{fig:examples}\end{figure}

We used the i$_{775}$ image, the deepest of the four images, to
search by eye to identify and study galaxies that have major axes
larger than 10 pixels (corresponding to 0.3 arcsec). The whole UDF
field was subdivided into 50 fields of 800x800 pixels for
identification of objects. Each of these subfields was further
subdivided and displayed in IRAF in 100x100 pixel increments,
which were scanned by eye to search for objects larger than 10
pixels (confirmed by measuring with contour plots out to the
2$\sigma$ limit).  Our full sample includes 884 galaxies.
Morphological classifications were made based on the i$_{775}$
images and aided by contour plots and radial profiles. The
galaxies were divided into 6 categories: chain (114 total),
clump-cluster (178), double (126), tadpole (97), spiral (269), and
elliptical (100). Figure \ref{fig:examples} shows 8 examples of
each type; the lines correspond to 0.5 arcsec.

Galaxy morphology can vary with wavelength, so we viewed many of
the catalogued objects at other ACS passbands and with NICMOS
(Thompson et al. 2005).  Generally the morphological
classification does not change significantly with wavelength
(e.g., Dickinson 2000) because it is based on only the most
fundamental galaxy characteristics, such as elongation and number
of giant clumps. Also, the NICMOS images have a factor of 3 lower
resolution, so they do not reveal the same fine structure as the
other images.

The distinguishing characteristics of the main types we classified
are:
\begin{itemize}
\item Chain: linear objects dominated by several giant clumps and
having no exponential light profiles or central red bulges.

\item Clump Cluster: oval or circular objects resembling chain
galaxies in their dominance by several giant clumps, and having no
exponential profiles or bulges.

\item Double-clump: systems dominated by two similar clumps with
no exponential profile or bulge.

\item Tadpole: systems dominated by a single clump that is
off-center from, or at the end of, a more diffuse linear emission.

\item Spiral: galaxies with exponential-like disks, evident spiral
structure if they have low-inclination, and usually a bulge or a
nucleus. Edge-on spirals have relatively flat emission from a
midplane, and often extended emission perpendicular to the
midplane, as well as a bulge.

\item Elliptical: centrally concentrated oval galaxies with no
obvious spiral structure.
\end{itemize}

Chain galaxies were first recognized by Cowie et al. (1995) using
the same definition as that here. Tadpole galaxies were defined by
van den Bergh et al. (1996) and examples from the UDF were
discussed by Straughn et al. (2004). Tadpole galaxies with short
tails were classified as ``comma'' type in the morphology review
by van den Bergh (2002). van den Bergh et al. (1996) also noted
objects like clump-clusters and called them proto-spirals.
Conselice et al. (2004) called these clump-dominated young disk
galaxies ``luminous diffuse objects,'' although some of their
sample included galaxies with bulges and exponential-like
profiles, unlike the clump-clusters here. Binary galaxies, like
our doubles, were noted by van den Bergh (2002). Other
classifications of galaxies in deep Hubble images, based on the
Hubble classification or DDO systems, were made by Brinchmann et
al. (1998), van den Bergh et al. (2000), and others, who generally
noted that an increasing fraction of distant objects fall outside
these conventional morphologies.

Distinctions between morphological types do not necessarily imply
there are significant physical differences. For example,
double-clump galaxies may be smaller versions of chain galaxies,
and the single-clump tadpoles may be smaller yet. The origins of
these types are not known. Also, chain galaxies and clump clusters
may be the same type of object viewed at different orientations
(Dalcanton \& Schectman 1996; O'Neil et al. 2000; Paper II).

\begin{figure}\epsscale{1.}\plotone{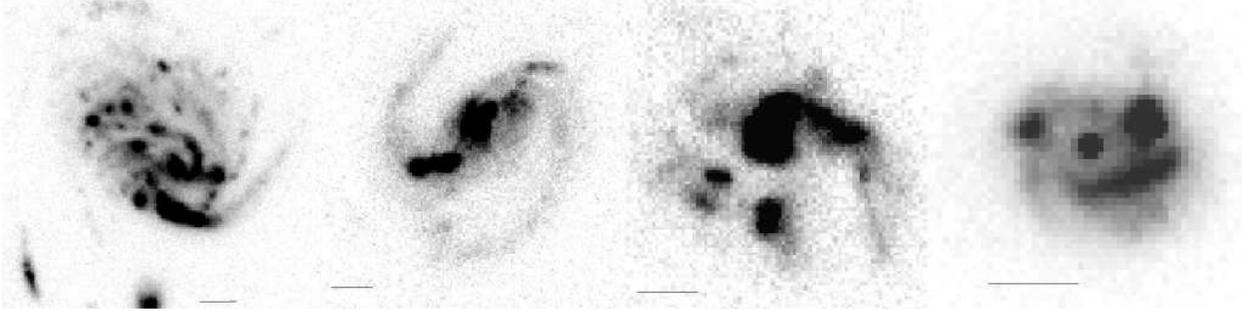}
\caption{Four spiral galaxies showing giant star formation
regions, tidal arms, and other asymmetries suggesting
interactions. UDF numbers are, from left to right: 8585, 2, 2993,
7112. UDF 2 has a bar.}\label{fig:4spirals}\end{figure}

Spiral galaxies are extremely varied. Many look highly disturbed,
as if by interactions with smaller companions. Figure
\ref{fig:4spirals} shows four more examples of spirals. The galaxy
on the left is very asymmetric and has a giant star formation
region along the bottom arm. The galaxy on the right is dominated
by two round clumps and an extended bright arm (``the cigar
smoker'').

Images of 10 different clump clusters from the UDF are in
Elmegreen \& Elmegreen (2005, hereafter Paper III), where their
clump properties were measured. Images of 8 different UDF
ellipticals are in Elmegreen, Elmegreen, \& Ferguson (2005,
hereafter Paper IV), which concentrated on those having giant blue
clumps in their centers (one of the ellipticals shown in Fig.
\ref{fig:examples}, UDF 4389, has blue clumps in its core, but
they cannot be seen in this image).

Ellipse fits were done on all of the spirals using IRAF to search
for radial variations in position angle and ellipticity. Bars have
a characteristic signature on such a plot as a result of the
twisting inner isophotes. The images were also viewed directly to
look for bars. In general, the most highly inclined galaxies do
not readily show bars even if they are present, and the short type
of bar that is commonly found in local late-type galaxies could
not be resolved.  Nevertheless, we found 26 barred spirals out of
269 spiral galaxies, a bar fraction of $\sim10$\%. This is a
factor of $\sim2$ smaller than in the Tadpole field, considering
even the inclination effects (Elmegreen, Elmegreen, \& Hirst
2004b, hereafter Paper V), and it is a factor of $\sim3$ smaller
than in the GOODS field (Jogee et al. 2004). The Tadpole field
contained 43 barred galaxies in a sample of 186 spirals and showed
clear correlations between apparent bar fraction and both
inclination and galaxy angular size. The GOODS field contained 80
bars out of 258 nearly face-on spirals. Both previous studies used
radial profiles of ellipticity and position angle to identify
bars, as we do here. Deeper fields like the UDF might be expected
to show fewer bars as physical resolution and surface brightness
dimming become more of a problem at high redshift.  Bar
destruction by severe interactions (Athanassoula \& Bosma 2003)
could play a role at high $z$ too. A more thorough study of the
UDF bars will be addressed in another paper. Early searches for
barred galaxies found relatively few examples (e.g., van den Bergh
et al. 2002), probably because of the lower angular resolution of
the WFPC2 camera. Sheth et al. (2003) was the first to suggest
that the relative numbers of bars might be the same as that found
locally.

\begin{figure}\epsscale{1}\plotone{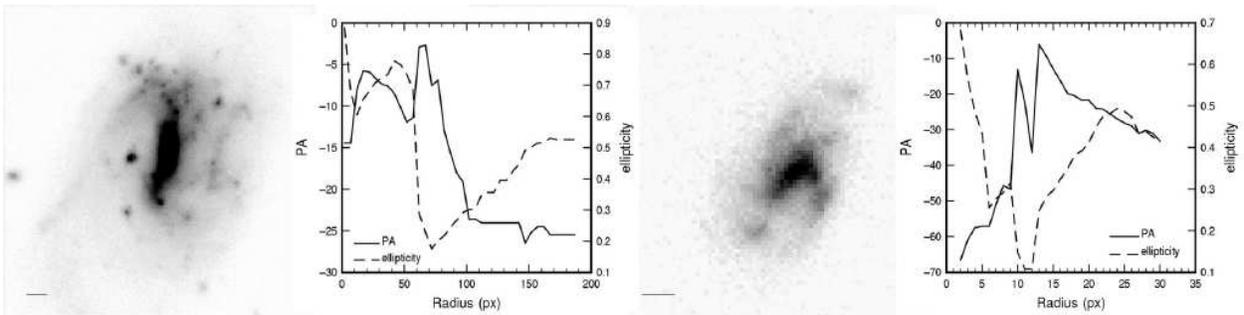}
\caption{The barred spirals UDF 1971 (left) and UDF 9341 are shown
with their radial distributions of position angle and
ellipticity.} \label{fig:bar}\end{figure}

Two examples of barred spirals in the UDF are shown in Figure
\ref{fig:bar}, along with the radial profiles of ellipticity and
position angle.  They are the galaxies UDF 1971 (left) and UDF
9341. The bars and spirals are more irregular than in local
galaxies, suggesting chaotic stellar orbits and poorly defined
resonances. The bars could therefore be young. Star formation is
enhanced in many small clusters near the top end of the bar in UGC
1971.

The distribution functions of the colors and magnitudes of the
clumps that have been measured in chains, clump-clusters, doubles,
and tadpoles, are all about the same (Paper I). Similar also are
the colors and magnitudes of clumps in the cores of 30 ellipticals
in the UDF (Paper IV).  These colors and magnitudes give clump
masses in the range from $10^6$ M$_\odot$ for the smallest
ellipticals (Paper IV) to $10^9$ M$_\odot$ for the best studied
clump-clusters (Paper III), based stellar population models from
Bruzual \& Charlot (2003). The clumps are clearly smaller and
bluer than the bulges of spiral galaxies (Paper I). The irregular
morphology of the first four types suggests a history dominated by
galaxy interactions, clump accretions, and large-scale
instabilities leading to star formation.  The highly clumpy
structure of the elliptical galaxies suggests a similar history of
clump accretions.

Most of the galaxies in this catalog are also listed in the UDF
catalog without morphological classifications (Beckwith et al.
2005, listed on the STScI website
\footnote{http://archive.stsci.edu/pub/hlsp/udf/acs-wfc/h\_udf\_wfc\_VI\_I\_cat.txt}).
Some of the galaxies here are not listed in the UDF catalog
because they appear near bright stars or near the edge of the
field of view. The parameters used for SExtractor were chosen to
optimize different aspects of the UDF than our morphological
system, so there is not a one-to-one correspondence.  Our catalog
has 9 chain galaxies, 14 clump-cluster galaxies, 10 double
galaxies, and 1 tadpole galaxy with multiple UDF numbers, and
there are 12 chains, 16 clump-clusters, 11 doubles, 8 tadpoles, 28
spirals and 6 ellipticals which are not in the UDF catalog at all.
One object, UDF 5898, is two separate spiral galaxies which
SExtractor called one, using the midpoint as a position; we call
these UDF 5898A and UDF 5898B. The unlisted and multiply-listed
galaxies amount to 154 objects, or 17\% of the total sample.

The integrated AB magnitudes were measured in all four filters
from counts within a rectangular box whose edges were defined by
the i$_{775}$ isophotal contours 2$\sigma$ above the sky noise
(corresponding to a surface brightness of 26.0 mag arcsec$^{-2}$).
The integrated magnitudes depend only weakly on the choice of
outer isophotal contour: 10-pixel extensions changed the
magnitudes of isolated objects by less than 0.1 mag. The
integrated i$_{775}$-z$_{850}$ colors of the galaxies range from
-0.7 to 1.3. The integrated i$_{775}$ magnitudes range from 18.5
to 29 mag, with surface brightnesses from 22.5 to 27 mag
arcsec$^{-2}$. The details of some clump properties are given in
separate studies of different subsamples (Papers III, IV).

Tables 1-6 list the galaxies by morphological type. In the first
part of each table, the galaxies with a single UDF number are
listed. In the second part, galaxies with multiple UDF numbers are
given along with the combined average surface brightnesses in
i$_{775}$, $\mu_{\rm I}$, the apparent magnitudes, i$_{775}$, and
the three colors. In the third part, galaxies with no UDF numbers
are given along with the coordinates of their centers and other
properties. We do not repeat the colors and magnitudes of the
single sources because these measurements are in the UDF catalog;
they are the same as what we measured independently to within 0.1
mag (which is the same as the measuring error for UDF magnitudes).
Barred galaxies are indicated with an asterisk in the spiral list.

\section{Photometric Results}

\begin{figure}\epsscale{0.8}\plotone{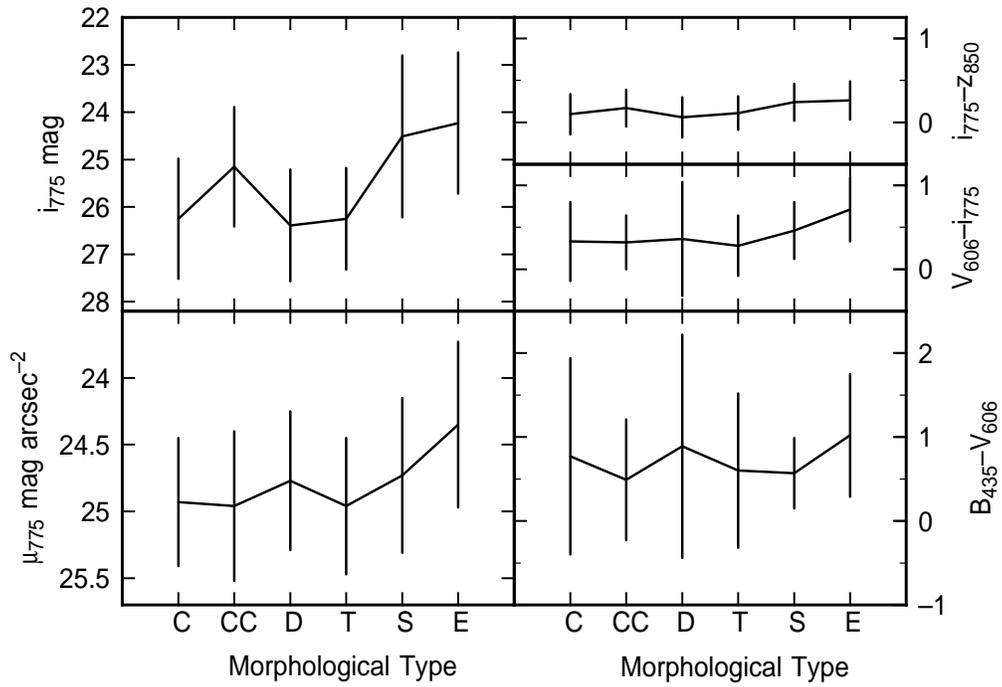}\caption{The
average magnitudes (upper left), surface brightnesses (lower
left), and colors (right) are shown for each galaxy type, with
$1\sigma$ rms fluctuations from sample varieties.  Ellipticals are
the reddest and brightest
galaxies.}\label{fig:colormagall}\end{figure}

The average color and magnitude values for each galaxy type are
shown in Figure \ref{fig:colormagall}. This plot has more
observational relevance than physical because the galaxies are a
mixture of redshifts and k-corrections. The average i$_{775}$
surface brightness is fairly constant at $\sim24.8$ mag
arcsec$^{-2}$ for all types except ellipticals, which are about
0.5 mag brighter. This constancy is likely the result of the
surface brightness limit of the survey ($2\sigma$ is 26.0 mag
arcsec$^{-2}$), with the averages being larger than this limit
because of central light concentration. The average i$_{775}$
integrated magnitude is about 26.3 for chains, doubles, and
tadpoles, 24.5 for spirals, 24.2 for ellipticals, and 25.2 for
clump clusters. The ellipticals have the reddest colors in all
indices.

The surface brightness distribution of all the galaxies peaks at
$\sim 25$ mag arcsec$^{-2}$, rapidly declines beyond the 2$\sigma$
limit of 26.0 mag arcsec$^{-2}$, and is virtually null beyond the
3$\sigma$ limit of 26.6 mag arcsec$^{-2}$ (see also Fig.
\ref{fig:wlvsmag} below).

Figure \ref{fig:overlay} plots B$_{435}$-V$_{606}$ versus
$i_{775}-z_{850}$ for all galaxies in our sample, divided
according to morphology. Note the Lyman Break galaxies of various
types, which have extremely red B$_{435}$-V$_{606}$ colors. Except
for these, the color distributions are similar with a large number
of galaxies at B$_{435}$-V$_{606}\sim0$ and a second diagonal
branch going up to B$_{435}$-V$_{606}$$\sim2$.  The curves
superposed on the data are Bruzual \& Charlot (2003) evolution
models that assume star formation began at $z=6$ for all cases and
then decayed exponentially over time with e-folding times of 0.01
Gy, 0.03 Gy, 0.1 Gy, 0.3 Gy, 1 Gy, 3 Gy, and infinity. Each curve
traces the sequence of decay times with the longest decay times
corresponding to the bluest colors. The different curves are for
different galaxy redshifts, from 0 to 4 in steps of 0.5 (see color
coding in figure caption). The models are described more in Paper
III.

The distribution of colors reveals several things about these
galaxies. The chains, clump-clusters, doubles, and tadpoles have
similar color distributions, and the spirals have about the same
distribution too, but the spirals do not have the very red
B$_{435}$-V$_{606}$ and i$_{775}$-z$_{850}$ components that are
present in the more irregular systems. These red points correspond
to the highest redshifts, $z\sim4$, so the difference in
distribution suggests the spirals do not generally go to as high a
redshift as the more irregular and clumpy galaxies. There is a
spiral, UDF 7036, that has the colors of a B-band drop-out,
however, so it could be at $z\sim4$ (see Sect. \ref{sect:lbg} and
Table \ref{tab:lbg}). Selection effects could remove the most
easily recognized spirals, which are the face-on spirals, from our
sample (see Sect. \ref{sect:axial}). The irregular and spiral
galaxies both populate the bluer regions in Figure
\ref{fig:overlay}, which suggests that both types linger around in
cosmological time after the spirals become obvious (see also
Franceschini et al. 1998).

The ellipticals also extend in Figure \ref{fig:overlay} to the
high V$_{606}$-i$_{775}$ and B$_{435}$-V$_{606}$ values of the
clumpy types. The ellipticals are slighlty redder than the other
types, lacking the big clump of points around
i$_{775}$-z$_{850}\sim0$ and B$_{435}$-V$_{606}\sim0$ that all the
others have.  Evidently the ellipticals have much less star
formation at the epoch of observation, as is well known (e.g.,
Ferguson, Dickinson \& Williams 2000).

\section{Relative Numbers}

Figure \ref{fig:fractions} plots the number and fraction of
galaxies of each type as a function of i$_{775}$ magnitude. The
counts of galaxies with spiral or elliptical morphologies peak at
about the same apparent magnitude, which is slightly brighter than
the peak of the clump clusters. The counts of chain, double, and
tadpole galaxies all peak at about one magnitude fainter. The
fractions follow a similar trend, with spirals dominating the
bright galaxies, while chains, doubles, and tadpoles dominate the
faint galaxies. As is well known, the galaxies with the most
unusual morphologies increase in dominance at the faintest
magnitudes, beyond i$_{775}\sim26$.

The trend for chains, doubles, and tadpoles to dominate at the
faintest magnitudes illustrates an important selection effect
discussed in the next section. We believe on the basis of this
trend, and on the basis of the distribution of axial ratios for
spirals, that the spirals begin to disappear at i$_{775}\sim26$
because the face-on spirals become too faint to see at the surface
brightness limit of the survey ($\sim26.0$ mag arcsec$^{-2}$ for
$2\sigma$ detections). Edge-on spirals have higher surface
brightnesses than face-on spirals, and are observed to fainter
magnitudes. In the distribution of axial ratios (see below), there
is a clear and unusual drop in the count for face-on spirals. In
the same way, the chain galaxies extend to fainter magnitudes than
the clump-clusters in Figure \ref{fig:fractions} because chain
galaxies are probably edge-on clump clusters (Paper II). Edge-on
galaxies tend to have higher surface brightnesses and fainter
apparent magnitudes than face-on galaxies (see models in Paper I).
Thus, the faintest galaxies at the surface-brightness limit of
detection are observed to be linear structures, many of which are
probably flattened objects viewed edge-on. The tadpole galaxies,
for example, could be strongly interacting galaxies with a
$10^9-10^{10}$ M$_\odot$ merger remnant seen as a single clump at
one end and a long tidal tail seen as the diffuse lane (e.g., see
Straughn et al. 2004).

\section{Axial Ratios}
\label{sect:axial}

Figure \ref{fig:ellfractwl} compares the distribution of axial
ratios, $W/L$, for elliptical galaxies in the UDF field with the
distribution for local ellipticals in the Third Reference
Catalogue of Bright Galaxies (de Vaucouleurs et al. 1991; RC3).
The UDF ratios came from the ellipse fits in the published UDF
catalog. The error bars equal $N^{1/2}$ for number $N$. For the
UDF distribution, the dashed histogram is for the half of the
sample with the brightest average surface brightness ($\mu_I<24.4$
mag arcsec$^{-1}$). This bright half is slightly rounder on
average than the faint half, as observed locally (Tremblay \&
Merritt 1996). The overall distribution for UDF ellipticals
resembles the distribution for local galaxies, particularly for
the high surface brightness ellipticals, suggesting that
elliptical galaxies relaxed to their current forms at very early
times.

The distribution of axial ratios for spiral galaxies is shown in
Figure \ref{fig:spiralwl}. The spiral galaxies in the deep field
background of the Tadpole Galaxy (Tran et al. 2003; Paper I) are
shown on the upper left.  The UDF spirals are on the bottom left.
We divided the UDF spirals into the faint surface-brightness half
and the bright surface-brightness half and plotted the separate
distributions on the right of Figure \ref{fig:spiralwl}.  For the
UDF spirals, the solid-line histograms use the ellipticities given
in the UDF catalog and the dashed-line histograms use
ellipticities derived individually for each galaxy from ellipse
fits in IRAF. Because the spirals are irregular, we did not want
to use the SExtractor ellipticities exclusively. To check the
first two ellipticity measures, we also determined the axial
ratios ($W/L=1-$ellipticity) by eye, based on the length ($L$) and
width ($W$) of the average 2-$\sigma$ contour limits for each
galaxy. All three axial ratio determinations correlated well with
each other, to within 0.1 in $W/L$. In what follows we consider
only the axial ratios determined by the UDF automatic ellipse fit
and by our IRAF ellipse fits, which are more objective than the
visual fits to contour shapes.

The distribution of axial ratios for spirals does not resemble the
local distributions for spirals, which is reproduced in Figure
\ref{fig:spiralwlrc3} using data from the RC3. The local
distribution is relatively flat from $W/L\sim1$ down to
$W/L\sim0.1$, depending on Hubble type, indicating that the
objects are round disks with intrinsic axial ratios comparable to
this lower limit. The only exception is for spirals of de
Vaucouleurs type 9, which are Sm irregulars and known to be
relatively thick.

The UDF and Tadpole-Field spirals have a distribution of axial
ratios that is peaked in the middle, which means that the catalog
lacks face-on examples ($W/L\sim1$). The fall-off at low $W/L$ is
probably from the intrinsic disk thickness, as for local galaxies,
although these spiral disks are evidently $\sim2$ times thicker
than the local versions (e.g., Reshetnikov et al. 2002; Papers II,
III). The fall-off at high $W/L$ could arise if the spirals are
intrinsically elongated disks, i.e., not circular, as suggested
for local irregular galaxies (Binggeli \& Popescu 1995; Sung et
al. 1998). Alternatively, they could be circular disks that tend
to be overlooked when viewed face-on. We consider these two
options in detail now.

Figure \ref{fig:wl} shows several models for the projected axial
ratios of intrinsically oval, i.e., triaxial, disks. The two on
the left show simple cases and the two on the right are sample
fits to the UDF or Tadpole data. The model on the bottom left is
for circular thick disks with a range of relative thicknesses,
$Z/L=0.225$ to 0.375, viewed in projection at random angles.  The
flat part extending to $W/L=1$ is a result of the circular form of
the disk itself, and the rise and fall below $W/L\sim Z/L$ is for
the edge-on case, when the minimum axial ratio is viewed. The top
left is for an ellipsoid with an axial ratio of 1:0.6:0.2. The
peak at $W/L=0.6$ is from the main disk axial ratio, when the disk
is viewed face-on. The two peaks at $W/L=0.2$ and 0.33 are for
edge-on versions; the lower ratio corresponds to the case where
the oval disk is viewed edge-on parallel to the minor axis
(showing the major axis length in projection), and the higher
ratio corresponds to the case where the oval disk is viewed
edge-on along the major axis (showing the minor axis length in
projection). Neither of these ideal cases matches the data, but
mixtures of them do. The models on the right span a range of
intrinsic $W/L$ and $Z/L$ ratios, viewed in projection for
orientations of the ellipsoid at random angles in 3D space.

Evidently, the distribution of axial ratios observed for the UDF
is consistent with intrinsically elongated spiral disk galaxies.
If the disks are actually distorted like this, then tidal forces
would be a likely explanation. Such an interpretation is plausible
in view of the importance of galaxy interactions at high redshift.
However, visual examination of each UDF spiral suggests that very
few of the elongated cases have companion galaxies or other direct
evidence of a strong interaction. Thus we consider another
explanation, as follows.

Recall that Figure \ref{fig:spiralwl} showed that the distribution
of axial ratios for the low surface brightness spirals is slightly
more skewed toward edge-on than the distribution for the high
surface brightness spirals. This distortion is shown again in
Figure \ref{fig:wlvsmag}, which plots the i$_{775}$ magnitudes and
average surface brightnesses (out to the $2\sigma$ contour) versus
the axial ratios. The solid red line is the average of the plotted
points, which are from the UDF catalog. The dashed red line is the
average for the W/L values we measured based on IRAF ellipse fits
(these values are not plotted). The density of points on this plot
increases for intermediate $W/L$, reflecting the peak at
intermediate $W/L$ in Figure \ref{fig:spiralwl}. The density of
points shifts to even lower $W/L$ at the faintest surface
brightnesses (top part of the top panel), and this shift also
parallels the trend in Figure \ref{fig:spiralwl}.

The bottom part of Figure \ref{fig:wlvsmag} shows that the average
flux gets $\sim2$ mag brighter as the spiral galaxies become more
face-on. This brightening also occurs for local spirals because
the total flux increases as the projected area increases and stars
become less obscured by midplane dust.

The top part of Figure \ref{fig:wlvsmag} indicates that the
average surface brightness is either constant or increases
slightly ($\sim0.5$ mag) for more face-on spirals, which is
contrary to expectations. Locally the surface brightness decreases
for face-on disks because the path length through the stars
decreases (Paper I).  The situation changes, however, if most of
the UDF surface brightnesses for spirals are below the detection
limit. This limit is 26 mag arcsec$^{-2}$ for the faintest contour
that defines our sample, so the average surface brightness of a
typical galaxy inside this contour is greater, perhaps by a
magnitude. The two dashed green lines in this figure represent the
approximate region where we begin to lose galaxies at the surface
brightness limit. Above $\sim26$ mag arcsec$^{-2}$, the integrated
surface brightness would be fainter than the lowest contour
considered here ($2\sigma$) and the galaxy would not likely be in
our survey. Thus we should ask what a distribution of surface
brightnesses should look like if the average surface brightness
decreases with increasing $W/L$ as for local galaxies, and as it
decreases, the detection limit is passed so the fainter galaxies
become lost in the noise. The solid green line shows what the
average distribution of surface brightnesses should be if a
face-on galaxy is at 25.9 mag arcsec$^{-2}$ (where $W/L=1$) and
edge-on galaxies get brighter in proportion to the path length
through the disk (i.e., $\mu=25.9-2.5\log L/W$).  This solid line
is well populated with observed galaxies at low $W/L$ so the
number of points in the plot is large there, but as $W/L$
increases, the surface brightness limit comes in and the number of
points centered on the line decreases.

Another way to see this surface brightness limit is with Figure
\ref{fig:fractions}. There we noted that the spiral count (in the
left panel) drops precipitously for galaxies fainter than
i$_{775}\sim25.5$ mag, which is about where the distributions of
chains, doubles, and tadpoles -- all linear systems -- begin to
peak. This drop may be combined with a model result in Paper I to
explain the factor-of-5 drop in the axial ratio distribution
function between $W/L=0.5$ and 1 (Fig. \ref{fig:spiralwl}). Paper
I showed in its Figure 8 a radiative transfer model through a
thick exponential disk with various levels of extinction; the
average disk surface brightness of the projected disk was plotted
versus the axial ratio.  As $W/L$ increased from 0.5 to 1, the
average surface brightness decreased by factors of 1.92, 1.78, and
1.74 for perpendicular extinctions to the midplane at the galaxy
center equal to 0.33, 1.33, and 1.77 optical depths. The limiting
case with no extinction would be a factor of 2, which is the
inverse ratio of the two $W/L$ values, as this intensity ratio is
mostly the ratio of path lengths along the line of sight.  For
these three extinctions, the surface brightness dimming in going
from $W/L=0.5$ to 1 corresponds to 0.71, 0.63, and 0.61 mag
arcsec$^{-2}$.  Such magnitude changes correspond, in Figure
\ref{fig:fractions}, to more than half of the full drop in spiral
galaxy counts (left-hand panel). This means that at the limit of
the UDF survey, projection effects alone can change a galaxy that
would be detected at $W/L<0.5$ to one that would be below the
limit of detection at $W/L\sim1$.

A simple model for the probability of detection of disk galaxies
near the surface brightness limit of a survey is shown in Figure
\ref{fig:faceon}.  This model assumes that the intrinsic surface
brightness distribution of face-on spiral galaxies, in magnitudes
arcsec$^{-2}$, is Gaussian, i.e., that the distribution of
intensity is log-normal. This assumption is consistent with the
narrow range of surface brightnesses for modern spiral galaxies
(Freeman 1970) and with the Tully \& Fisher (1977) relation, which
suggests the same narrow range. In terms of the model on the top
of Figure \ref{fig:wlvsmag}, the breadth of the Gaussian would
give the intrinsic vertical range of points around the solid green
line. For Figure \ref{fig:faceon}, we integrate over this
log-normal distribution to find the relative number of galaxies
detected above some minimum intensity level. The absolute value of
the intensity at the peak of the Gaussian is arbitrary so we take
it to be $I_f({\rm peak})=1$, and we take the Gaussian dispersion
to be $\sigma$. The observed intensity of a galaxy with face-on
intensity $I_f$ is assumed to be $I_fL/W$, which is approximately
the case without extinction (for $W/L$ not too small).  This means
that if an observed intensity is $I$, then the intrinsic face-on
intensity for use in the distribution function is $IW/L$. Thus we
plot the detection probability:
\begin{equation}
P_{\rm detect}={{\int_{\ln I_{\rm thresh}}^\infty
\exp\left(-0.5\left[\ln\left\{IW/L\right\}/\sigma\right]^2\right)
d\ln I} \over{\int_{-\infty}^\infty
\exp\left(-0.5\left[\ln\left\{IW/L\right\}/\sigma\right]^2\right)
d\ln I}}. \label{eq:p}
\end{equation}
The intensity threshold $I_{\rm thresh}$ and intrinsic dispersion
were chosen to match the observed fall-off in the distribution of
axial ratios between $W/L=0.5$ and 1. This gives $\ln I_{\rm
thresh}$ between 0.25 and 0.5, or $I_{\rm thresh}$ between
$e^{0.25}=1.3$ and $e^{0.5}=1.6$ times the intensity at the peak
of the log-normal distribution for the two cases plotted. The fit
also gives $\sigma=0.2$.

The fits in Figure \ref{fig:faceon} indicate that the fall-off in
spiral galaxy counts between $W/L=0.5$ and 1 corresponds to a
most-probable, face-on, galaxy surface brightness that is fainter
than the threshold of detection by a factor between 1.3 and 1.6.
Thus a high fraction of face-on spiral galaxies are likely to be
missed and high fraction of inclined spiral galaxies are likely to
be seen (if they satisfy our $>10$ pixel size selection).

The fall-off in galaxy count for face-on spirals does not appear
for local galaxies (Fig. \ref{fig:spiralwlrc3}).  This is because
there is a limited range in surface brightness for most spiral
galaxies and the imaging surveys that made the RC3 went
considerably deeper than the average galaxy surface brightness.
The fact that there is also not a large increase in local counts
at small $W/L$ indicates that there is not a relatively large
number of thin, ultra-low surface brightness galaxies in the local
Universe.  The situation is apparently different in the UDF
because even though galaxies tend to be intrinsically more intense
than they are locally, i.e., because of their higher star
formation rates, cosmological surface brightness dimming, which is
proportional to $(1+z)^4$, knocks down their apparent surface
brightnesses by large factors, corresponding to 3, 4.8, 6, and 7
magnitudes arcsec$^{-2}$ for $z=1$, 2, 3 and 4. Thus only the
brightest of the local galaxies, viewed at these redshifts, would
be detected in the UDF survey, and most would be lost unless they
were viewed nearly edge-on.

The distribution of axial ratios for chain and clump cluster
galaxies in the UDF was shown in Paper III. It is flatter than the
spiral galaxy distribution, perhaps because the clumps in clump
clusters are very bright and they can still be seen in the face-on
versions even if the interclump medium is close to or below the
survey limit.

Figure \ref{fig:fractions} showed a large difference in the
apparent magnitude distribution for spirals compared to the linear
systems (chains, doubles and tadpoles) and a modest difference
compared to clump clusters, which may be face-on versions of
chains. The linear systems dominate at the faintest magnitudes and
some of the reason for this may be the surface brightness
selection effects just described, especially since the spiral
sample in Figure 6 contains the full range of axial ratios.
However, not all of the difference in these distributions is
likely to result from surface brightness selection. Paper I showed
that even the edge-on spirals in the Tadpole field, as identified
by their bulges and exponential disks, tend to have brighter
apparent magnitudes and slightly brighter surface brightnesses
than the chains, doubles, and tadpoles.  This is true also for the
edge-on spirals in the UDF sample (not shown here). This
difference among the purely edge-on systems could be the result of
a smaller average redshift for the spirals than for the chains,
doubles and tadpoles, which would be consistent with evolution
over time from irregular structures to smooth spiral disks.  For
example, we noted in Paper I that the spirals in the Tadpole field
tend to have slightly larger angular sizes than the doubles and
tadpoles (by 50\% on average). The redshift distributions of
galaxies with various morphologies will have to be studied in
order to determine the relative importance of surface brightness
selection and evolutionary effects.

\section{Lyman Break Galaxies}
\label{sect:lbg} Steidel et al. (1999 and references therein)
identified high z galaxies photometrically by their redshifted
Lyman break. Adelberger et al. (2004) and others applied the
technique to galaxies between redshifts 1 and 3. In the UDF field,
Lyman Break Galaxies (LBGs) appear as B-band drop-outs at $z\sim4$
and V-band drop-outs at $z\sim5$. Giavalisco et al. (2004) gave
color criteria for Lyman Break galaxies, based on a sample of LBGs
in the GOODS/ACS survey. Based on their criteria, we find in our
UDF sample 30 B-band drop-outs (10 chains, 4 clump clusters, 8
doubles, 4 tadpoles, 1 spiral, and 3 ellipticals) and 13 V-band
drop-outs (1 chain, 3 clump-clusters, 8 doubles, and 1 tadpole).
These galaxies are identified by their type and UDF catalog number
in Table \ref{tab:lbg}. In Figure \ref{fig:overlay}, many of the
objects redder than $\sim1.5$ in B$_{435}$-V$_{606}$ are B-band
drop-outs. There are many other LBG galaxies in the UDF that are
smaller than our 10 pixel major axis limit (Bouwens et al. 2004).
Bunker et al. (2004) and Yan \& Windhorst (2004) reported over 100
i-band drop-out ($z\sim6$) galaxies in the UDF; one of our V-band
drop-out chain galaxies was also reported at redshift z=5.4 by
Rhoads et al. (2004).

Most of the drop-outs in Table \ref{tab:lbg} are linear galaxies
(doubles, chains, tadpoles) and relatively few are disks (spiral).
This is consistent with the common perception that disks form
late, but it may also contain a selection bias like that discussed
in the previous section if most face-on disks are fainter than the
survey limit.

Madau et al. (1998) determined star formation rates (SFRs) from
models of star clusters with Salpeter initial mass functions based
on observed UV fluxes. Using their conversions, and assuming that
the rest wavelength of 1500 \AA\ is shifted approximately to
i$_{775}$-band for the $z\sim4$ galaxies and to z$_{850}$ band for
$z=5$, we calculated the SFRs (in M$_\odot$/year) of our LBGs. The
results are listed in Table \ref{tab:lbg}.  The star formation
rates for the V-band drop-outs in this sample tend to be larger
than for the B-band drop-outs. In comparison, Immeli et al. (2004)
predicted a SFR of about 20 M$_\odot$ yr$^{-1}$ from their chain
galaxy models, which assumed galaxy sizes a factor of five larger
than those of our LBGs.  Somerville, Primack, \& Faber (2001) got
rates of $\sim1-30$ M$_\odot$ yr$^{-1}$ for the largest galaxies
considered in their Lyman Break models.

\section{Summary}

We classified and measured 884 galaxies larger than 10 pixels in
the Hubble UDF.  Six prominent types were used: chain,
clump-cluster, double, tadpole, spiral, and elliptical. Their
colors, magnitudes, and surface brightnesses are similar,
reflecting their likely formation over a large range of times. The
distribution of axial ratios for elliptical galaxies is similar to
that in the local field, suggesting that the elliptical shapes
were quickly established and remained unchanged over time. The
distribution of axial ratios for spiral galaxies was distinctly
different than for local galaxies with a deficit of face-on
spirals in the deep field. Models illustrate how this could be the
result of an intrinsically oval shape, as might be expected for
highly perturbed galaxies. However, other models suggest that most
of the face-on spirals could be lost below the surface brightness
limit of the survey, while the more highly inclined spirals could
still be seen. A significant lack of spiral galaxies with low
axial ratios implies an intrinsic disk thickness of 0.2 to 0.3
times the disk length.  This implies the spiral disks are thicker
than local spirals by a factor of $\sim2$ to 3.

Twenty-six barred spiral galaxies were found, a fraction of the
total spirals equal to about 10\%. This is a factor of 2 less than
other deep field studies but not inconsistent with them
considering projection effects and surface brightness dimming. In
addition, bars become harder to distinguish at high redshifts
because the spiral galaxies generally get more irregular.

We thank the Vassar College URSI program and the Keck Northeast
Astronomy Consortium exchange program for undergraduate research
support for M.S. and D.S.R.;  B.G.E. is supported by the National
Science Foundation through grant AST-0205097.

\newpage
\begin{table}
\begin{center}
\caption{Chain Galaxies\label{tab:chain}}
\begin{tabular}{cccccccccccccc}
\tableline\tableline
UDF &&&&&&&&&&&&&\\
\tableline
    65&    98&   133&   169&   170&   173&   401&   412&   521&   666&  1236&  1287&  1383&  1428\\
  1531&  1678&  2034&  2265&  2394&  2426&  2750&  2763&  2764&  2998&  3001&  3091&  3143&  3214\\
  3299&  3354&  3368&  3459&  3529&  3607&  3693&  3779&  3974&  3977&  4010&  4013&  4243&  4387\\
  4411&  4568&  4581&  4660&  4674&  5010&  5014&  5053&  5225&  5380&  5454&  5497&  5787&  6017\\
  6278&  6391&  6450&  6478&  6523&  6539&  6607&  6634&  6709&  6715&  6922&  6941&  7037&  7076\\
  7269&  7352&  7617&  7681&  7737&  7816&  7868&  8092&  8097&  8268&  8372&  8624&  8681&  8731\\
  8801&  8805&  9171&  9270&  9414&  9676&  9839&  9861&  9974&     &     &     &    &     \\
  \end{tabular}
\begin{tabular}{lccccccccc}
\tableline\tableline
UDF\tablenotemark{a}&&&&&$\mu_{\rm I}$&I&B-V&V-i&i-z\\
\tableline
1309+1347 &&&&&      24.29 & 23.85 &0.55&0.71&0.45\\
1419+1396 &&&&&       25.20 &25.77 &0.61&0.05&-0.02\\
2436+2484 &&&&&      24.73& 25.74 &0.38&0.02 &0.01\\
3178+3194 &&&&&       25.88& 26.09 &0.23&0.09 &0.08\\
3458+3418 &&&&&      25.07& 26.04& 2.82& 0.63&0.08\\
4976+4981 &&&&& 24.47& 24.21 &0.04&  0.05 &0.29\\
8330+8369+46 &&&&&    25.46& 24.96 &0.80&0.06 & -0.08\\
9085+8865    &&&&&    24.61& 26.37 &1.99& 0.47&0.04\\
9934+... &&&&&  25.06 & 25.93& 0.09& 0.46&0.21\\
\tableline\tableline
Number&x&y&RA(2000)&DEC(2000)&$\mu_{\rm I}$&I&B-V&V-i&i-z\\
\tableline
C1&     3081.5 &3433&   3 32 43.9064&   -27 48 23.598& 24.32 & 24.79& 0.21&-0.09&0.02\\
C2&     6901.5 &2194.5& 3 32 35.2695&   -27 49 00.793& 24.47& 25.55 &0.88&0.24&0.10\\
C3&    5088   &1935    &3 32 39.3706&   -27 49 08.563& 24.60& 26.64 &2.26 &0.76&0.46\\
C4&     1602.4 &4288.1& 3 32 47.2496&   -27 47 57.921& 24.79& 26.03 &1.49 &  0.30 & -0.04\\
C5&     2294.5 &7179  & 3 32 45.6812&   -27 46 31.206& 24.83& 26.68& -0.01&  0.05 &  0.49\\
C6 &    7719   &2828.5& 3 32 33.4207&   -27 48 41.777& 24.87 & 24.29& 0.28& 0.62&0.27\\
C7 &    5270   &5057.5& 3 32 38.9568&   -27 47 34.890& 25.02& 26.71 & 1.90& 0.28&-0.12\\
C8 &    4276   &8312.5& 3 32 41.2011&   -27 45 57.229& 25.03& 28.40 & 1.13& 0.36&0.038\\
C9 &    6147   &5311.5& 3 32 36.9739&   -27 47 27.278& 25.11& 26.47 &1.23 &1.11 &0.65\\
C10&    5076.4 &9884.9& 3 32 39.3908&    -27 45 10.066& 25.24& 25.79& -0.16&0.43&0.27\\
C11&    5981.5 &2948.5& 3 32 37.3495&    -27 48 38.166& 25.26& 27.34& 0.16&0.09&       0.04\\
C12&    5966 &  2989  & 3 32 37.3845&    -27 48 36.951& 25.36& 27.60& 0.01&0.17&0.21\\
\tableline
\end{tabular}
\tablenotetext{a}{UDF 9934 also includes UDF 9912 and UDF 9949.}
\end{center}
\end{table}

\newpage
\begin{table}
\begin{center}
\caption{Clump Cluster Galaxies\label{tab:cc}}
\begin{tabular}{cccccccccccccc}
\tableline\tableline
UDF &&&&&&&&&&&&&\\
\tableline
    40&    84&    97&   126&   168&   187&   533&   551&   566&   769&   791&   797&   853&   857\\
   918&   941&  1041&  1112&  1266&  1269&  1316&  1362&  1375&  1589&  1666&  1681&  1693&  1775\\
  1859&  2012&  2174&  2240&  2291&  2340&  2350&  2350&  2463&  2499&  2538&  2853&  3021&  3031\\
  3037&  3154&  3243&  3270&  3460&  3483&  3688&  3689&  3745&  3752&  3778&  3799&  3844&  3877\\
  3881&  4006&  4084&  4093&  4094&  4253&  4262&  4370&  4616&  4807&  4860&  4999&  5050&  5107\\
  5136&  5190&  5201&  5216&  5491&  5501&  5548&  5579&  5620&  5634&  5683&  5685&  5748&  5827\\
  5837&  5860&  5878&  5896&  5946&  6056&  6133&  6187&  6394&  6396&  6438&  6456&  6486&  6499\\
  6645&  6710&  6785&  6821&  6837&  6854&  6939&  6943&  7185&  7227&  7230&  7328&  7432&  7526\\
  7559&  7647&  7678&  7725&  7756&  7786&  7905&  7995&  8022&  8042&  8125&  8217&  8262&  8270\\
  8314&  8461&  8551&  8682&  8710&  8749&  8765&  8802&  8859&  8880&  9000&  9112&  9159&  9273\\
  9299&  9347&  9356&  9396&  9409&  9474&  9837&  9853&     &     &     &     &     &     \\
  \end{tabular}
\begin{tabular}{lccccccccc}
\tableline\tableline
UDF\tablenotemark{a}&&&&&$\mu_{\rm I}$&I&B-V&V-i&i-z\\
\tableline
82+99+103&  &&&&    25.07&  23.59& 0.43&0.25&0.30\\
3034+3129&  &&&&     24.6 &   24.03& 0.82&0.18&0.08\\
3465+3597&  &&&&     24.32 &  23.48&  0.11&  0.04& 0.02\\
4245+...&    &&&&     26.16 &  25.08& 0.08& 0.25&0.32\\
4258+4458&  &&&&     25.12 &  25.56& 0.60&0.27&0.10\\
4765+4795&  &&&&     24.81 &  25.56& 2.03& 0.38&-0.01\\
4801+5271&   &&&&     25.17 &  23.87& 0.22&0.37&0.56\\
6235+6153&  &&&&     25.14 &  25.59& 0.01&0.17&0.33\\
6462+6886&  &&&&     25.51 &  23.48& 0.18&0.32&0.27\\
6713+6738&  &&&&     25.24 &  26.35& 3.86& 0.62&-0.08\\
7008+7045&  &&&&     25.35 &  26.16& 1.31& 0.32&0.08\\
7081+...& &&&&     25.86 &  24.52& 0.06&  0.105&0.34\\
9169+9332&  &&&&     24.88 &  24.48& 0.16&0.18&0.30\\
9166+9102&  &&&&     25.13 &  24.35& 0.32&0.63&0.62\\
\tableline
\end{tabular}
\tablenotetext{a}{UDF 4245 also includes UDF numbers 4450, 4466,
4501, and 4595. UDF 7081 also includes UDF 7136 and UDF 7170.}
\end{center}

\newpage
\begin{center}
\begin{tabular}{cccccccccccccc}
\tableline\tableline
Number&x&y&RA(2000)&DEC(2000)&$\mu_{\rm I}$&I&B-V&V-i&i-z\\
\tableline
CC1 &      4446  &1030&   3 32 40.8232 &   -27 49 35.706 &24.32 &24.11 &1.20 &0.24&0.05\\
CC2 &      3809.2& 1717.2& 3 32 42.2627 &    -27 49 15.082 &25.16&  24.52& 0.88&0.12&0.21\\
CC3 &     6154.5 &2458   &3 32 36.9586  &  -27 48 52.883 &25.14  &25.90 &0.056& 0.45&0.09\\
CC4 &     688    &5339   &3 32 49.3154  &  -27 47 26.377 &23.86  &22.00 &0.44&0.16&-0.07\\
CC5 &      769.3 & 4832.3& 3 32 49.1323 &   -27 47 41.579& 25.36 & 25.23& 0.02& 0.26&0.08\\
CC6 &      1117  & 5932.5& 3 32 48.3447 &   -27 47 08.580& 25.13 &26.14 &0.81&0.04&-0.21\\
CC7 &      1168.7& 6154.3& 3 32 48.2275 &   -27 47 01.927& 24.82 & 26.47& 0.49&0.04& -0.18\\
CC8 &      8296.5& 6395  & 3 32 32.1142 &   -27 46 54.784& 24.15 & 25.13& 0.99&0.24&0.03\\
CC9 &      9018  & 6086  & 3 32 30.4832 &   -27 47 04.055& 24.69 & 26.20& 0.17&0.04&0.02\\
CC10&      8824  & 6572  & 3 32 30.9217 &   -27 46 49.475& 24.62 & 24.18& 0.75&0.87& 0.72\\
CC11&      3702.4& 8496.8& 3 32 42.4974 &   -27 45 51.693& 23.63 & 24.38& 1.73& 0.30&-0.03\\
CC12&      4211.6&8945.3 &3 32 41.3461  &  -27 45 38.245 &24.07  &23.49 &0.37 &0.60&0.43\\
CC13&      4256.1 &9120.8& 3 32 41.2453 &   -27 45 32.980& 24.83 & 24.30& 0.02&0.10&0.07\\
CC14&      4385.7 &9269.3& 3 32 40.9523 &   -27 45 28.527& 24.03 & 23.17& 0.40&0.60&0.08\\
CC15&      4982   &9541  & 3 32 39.6044 &   -27 45 20.382& 24.53 & 23.68& 0.03&0.44&0.22\\
CC16&      6058.3 &9293.7& 3 32 37.1721 &   -27 45 27.811& 25.36 & 26.53& 2.32& 0.85&0.09\\
\tableline
\end{tabular}
\end{center}
\end{table}

\newpage

\begin{table}
\begin{center}
\caption{Double Galaxies\label{tab:doubles}}
\begin{tabular}{cccccccccccccc}
\tableline\tableline
UDF &&&&&&&&&&&&&\\
\tableline
    86&   178&   207&   209&   475&   637&   966&   985&  1000&  1416&  1619&  1662&  1667&  1796\\
  1955&  1992&  2032&  2215&  2461&  2558&  2943&  2954&  3248&  3351&  3373&  3454&  3513&  3527\\
  3701&  3895&  3907&  3933&  3967&  4040&  4072&  4082&  4097&  4137&  4228&  4313&  4461&  4479\\
  4569&  4581&  4603&  4644&  4685&  4686&  4699&  5098&  5159&  5251&  5304&  5405&  5456&  5574\\
  5632&  5669&  5784&  5788&  5928&  5956&  5962&  5964&  5983&  6128&  6137&  6139&  6209&  6254\\
  6411&  6487&  6543&  6681&  6754&  6808&  6957&  7159&  7234&  7318&  7622&  7711&  7815&  7907\\
  8207&  8273&  8326&  8327&  8419&  8637&  8664&  8878&  9024&  9092&  9110&  9139&  9304&  9308\\
  9330&  9406&  9414&  9505&  9706&  9720&  9777&     &     &     &     &     &     &     \\
\end{tabular}
\begin{tabular}{lccccccccc}
\tableline\tableline
UDF&&&&&$\mu_{\rm I}$&I&B-V&V-i&i-z\\
\tableline
1829+1752&&&&&  23.42&  24.52&  0.03&   0.20&   0.45\\
2061+2015&&&&&  23.95&  27.65&  -0.09&  0.13&   0.32\\
2130+2156&&&&&  25.01&  26.56&  0.00&   -0.18&  -0.11\\
3377+3398&&&&&  25.85&  27.69&  0.86&   5.27&   1.14\\
5326+5312&&&&&  25.21&  27.52&  0.11&   -0.09&  -0.21\\
5339+5322&&&&&  25.80&  26.83&  0.91&   0.12&   -0.12\\
6356+6351&&&&&  25.24&  27.34&  0.35&   -0.05&  -0.09\\
7694+7702&&&&&  24.82&  25.17&  1.51&   0.37&   -0.01\\
7812+7934&&&&&  26.08&  27.25&  0.93&   0.29&   -0.02\\
8966+9011&&&&&  24.70&  25.12&  0.74&   0.20&   0.11\\
\tableline\tableline
Number&x&y&RA(2000)&DEC(2000)&$\mu_{\rm I}$&I&B-V&V-i&i-z\\
\tableline
D1&             7302.1& 2360.1& 3 32 34.3636&    -27 48 55.827& 23.17&  24.76&  3.05&   0.92&   0.14\\
D2&             7409.8& 2568.1& 3 32 34.1200&    -27 48 49.588& 24.01&  23.92&  0.29&   0.08&   0.01\\
D3&             606.9&  5159.6& 3 32 49.4990&    -27 47 31.757& 24.78&  25.82&  0.96&   0.26&   -0.18\\
D4&             1760.7& 6704.9& 3 32 46.8884&    -27 46 45.420& 24.22&  25.57&  0.50&   -0.07&  -0.15\\
D5&             6490.2& 7924.1& 3 32 36.1967&    -27 46 08.902& 24.65&  25.65&  0.34&   0.13&   -0.02\\
D6&             8126.9& 7044.6& 3 32 32.4974&    -27 46 35.295& 24.80&  26.98&  0.97&   0.40&   -0.09\\
D7&             8730.4& 6496.7& 3 32 31.1333&    -27 46 51.734& 24.97&  26.74&  0.17&   -0.00&  -0.16\\
D8&             4036.3& 8687.9& 3 32 41.7425&    -27 45 45.965& 24.84&  26.11&  0.50&   0.08&   -0.01\\
D9&             4696.6& 564.7&  3 32 40.2569&    -27 49 49.668& 24.21&  25.19&  0.53&   0.38&   0.32\\
D10&            5211.6& 683.6&  3 32 39.0920&    -27 49 46.107& 24.70&  25.81&  0.36&   0.21&   0.15\\
D11&            1094.8& 6186.3& 3 32 48.3945&    -27 47 00.966& 24.86&  27.70&  0.88&   2.43&   -0.05\\
\tableline
\end{tabular}
\end{center}
\end{table}
\newpage
\begin{table}
\begin{center}
\caption{Tadpole Galaxies\label{tab:tadpoles}}
\begin{tabular}{cccccccccccccc}
\tableline\tableline
UDF &&&&&&&&&&&&&\\
\tableline
    38&    83&   110&   141&   176&   197&   285&   648&   719&   741&  1454&  1469&  1720&  2044\\
  2230&  2445&  2542&  2785&  2872&  2879&  2881&  3058&  3059&  3128&  3147&  3185&  3442&  3468\\
  3508&  3573&  3583&  3682&  3819&  3823&  3893&  3906&  3945&  4053&  4184&  4187&  4287&  4390\\
  4548&  4592&  4601&  4638&  4668&  4717&  4838&  4841&  4908&  4927&  5063&  5115&  5124&  5358\\
  5377&  5386&  5388&  5443&  5533&  5674&  5843&  6119&  6290&  6782&  6891&  6999&  7044&  7146\\
  7202&  7210&  7862&  7900&  7962&  8006&  8150&  8523&  8561&  8580&  8614&  8750&  8950&  8964\\
  9310&  9348&  9543&  9848&     &     &     &     &     &     &     &     &     &     \\

  \end{tabular}
\begin{tabular}{lccccccccc}
\tableline\tableline
UDF&&&&&$\mu_{\rm I}$&I&B-V&V-i&i-z\\
\tableline
4348+4352  &&&&&          26.28 &  26.34 & 1.20 & 0.15& 0.36\\
\tableline\tableline
Number&x&y&RA(2000)&DEC(2000)&$\mu_{\rm I}$&I&B-V&V-i&i-z\\
\tableline
T1 &   679.05&  5958.7&  3 32 49.3347   & -27 47 07.786 &25.11& 25.71 & 0.31&-0.06 &-0.07\\
T2 &   4170.6&  6467.2&  3 32 41.4410   & -27 46 52.587 &24.78&  25.17& 0.72&0.45&0.25\\
T3 &   3557.2&  8235.3&  3 32 42.8258   & -27 45 59.536 &24.74&  24.13& 0.62&0.73&0.19\\
T4 &   4290.2&  9000.2&  3 32 41.1684   & -27 45 36.599 &24.13 & 26.04& -0.02&0.56& 0.38\\
T5 &   3997.1&  8840.1&  3 32 41.8310   & -27 45 41.398 &24.47 & 25.92& -0.02&0.13& 0.35\\
T6 &   8623.6&  3759.6&  3 32 31.3751   & -27 48 13.847 &24.74 & 26.31& 3.11 &2.20& 0.52\\
T7 &   5885.9&  1300.1&  3 32 37.5667   & -27 49 27.618 &24.58 & 25.23& 1.72 &0.57& 0.08\\
T8 &   6236.8&  1462.7&  3 32 36.7730   & -27 49 22.742 &24.70 & 24.98& 0.24&0.49& 0.08\\
 \tableline
\end{tabular}
\end{center}
\end{table}

\newpage
\begin{table}
\begin{center}
\caption{Spiral Galaxies\tablenotemark{a}\label{tab:spirals}}
\begin{tabular}{cccccccccccccc}
\tableline\tableline
UDF &&&&&&&&&&&&&\\
\tableline
      1&     2&    13&    14&    15&    17&    20&    37&    43&    50&    55&    95&   131&   163\\
   192&   213&   235&   237&   295&   328&   355&   423&   424&   428&   446&   501&   503&   522\\
   542&   656&   662&   679&   697&   735&   833&   898&   916&   968&  1006&  1049&  1052&  1057\\
  1270&  1403&  1421&  1426&  1478&  1551&  1571&  1592&  1612&  1626&  1668&  1732&  1739&  1789\\
  1809&  1830&  1889&  1898&  1904&  1905&  1922&  1971*&  2017&  2104&  2170&  2189&  2306&  2332\\
  2333&  2358&  2462&  2525&  2607&  2652&  2753&  2799&  2813&  2993&  3006&  3013&  3068&  3075\\
  3097&  3101&  3123&  3180&  3203*&  3247&  3257&  3268&  3315&  3319&  3349*&  3372&  3422&  3445\\
  3472&  3492*&  3610&  3613*&  3680&  3785&  3822*&  3823&  3840&  3871&  4052&  4058&  4192&  4225\\
  4315&  4321&  4360&  4394&  4407&  4410&  4438&  4478&  4491&  4591&  4661&  4662&  4767&  4816\\
  4835&  4878&  4907&  4929&  4950&  5022&  5177*&  5268&  5276&  5286&  5365*&  5408&  5411&  5417\\
  5435&  5473&  5505&  5540&  5569&  5606*&  5614&  5615&  5658&  5670&  5694&  5697&  5753&  5805\\
  5812&  5864&  5898A\tablenotemark{b}&  5898B\tablenotemark{b}&  5922&  5989&  5995&  5999&  6008&  6038&  6051*&  6060&  6079&  6082\\
  6107&  6143&  6188&  6203&  6206&  6639&  6674&  6680&  6810&  6834&  6862*&  6911&  6933*&  6974\\
  6997*&  7022&  7036&  7067&  7071&  7112&  7203*&  7406&  7452*&  7495&  7556&  7664&  7688*&  7705\\
  7743&  7775&  7847&  7974&  8015&  8026*&  8040&  8049*&  8051&  8156&  8181&  8255&  8257&  8259\\
  8261&  8275&  8351&  8409&  8454&  8576&  8585&  8603&  8629&  8693&  8757&  8782&  8918&  9018*\\
  9074&  9125&  9183&  9204&  9253&  9341*&  9425&  9444*&  9455&  9599&  9609&  9649&  9724&  9759\\
  9807&  9834&  9895&     &     &     &     &     &     &     &     &     &     &     \\
  \tableline
\end{tabular}
\end{center}
\newpage
\begin{center}
\begin{tabular}{lccccccccc}
\tableline\tableline
Number&x&y&RA(2000)&DEC(2000)&$\mu_{\rm I}$&I&B-V&V-i&i-z\\
\tableline
s1* &  5102.6& 538.9 & 3 32 39.3387 &   -27 49 50.446 &24.54  &23.59 & 0.78   &0.61   &0.19\\
s2 &  6144.9& 1292.1& 3 32 36.9810 &   -27 49 27.860 &24.50  &24.92 & 0.18   &0.59   &0.17\\
s3 &  6211.3& 1497.7& 3 32 36.8307 &   -27 49 21.692 &25.29  &25.56 & 0.21   &0.71   &0.31\\
s4 &  2806.1& 2886.6& 3 32 44.5297 &   -27 48 39.986 &24.17  &24.43 & 0.88   &0.28   &0.21\\
s5 &  3916.9& 2413.6& 3 32 42.0184 &   -27 48 54.192 &25.35  &25.63 & 1.25   &0.92   &0.16\\
s6 &  5150.1& 2557.2& 3 32 39.2298 &   -27 48 49.898 &24.73  &25.63 & 0.67   &0.09   &0.04\\
s7 &  7100.9& 2372.6& 3 32 34.8186 &   -27 48 55.451 &24.55  &24.48 & 0.76   &0.20   &0.10\\
s8 &  7845.6& 3062.1& 3 32 33.1344 &   -27 48 34.769 &24.82  &23.85 & 0.58   &0.65   &0.14\\
s9 &  1982.1& 3846.9& 3 32 46.3917 &   -27 48 11.164 &25.58  &25.72 & 0.20   &0.01   &-0.09\\
s10&  1958.1& 3902.1& 3 32 46.4459 &   -27 48 09.507 &24.96  &25.09 & 1.14   &0.60   &0.09\\
s11&  1640.9& 4158.1& 3 32 47.1627 &   -27 48 01.822 &25.03  &25.36 & 1.05   &0.26   &-0.01\\
s12&  8265.1& 3406.9& 3 32 32.1858 &   -27 48 24.427 &25.19  &23.60 & 0.91   &0.23   &0.18\\
s13*&  9257.4& 4315.4& 3 32 29.9420 &   -27 47 57.173 &23.96  &22.68 & 1.84   &1.21   &0.48\\
s14&  9193.8& 4319.4& 3 32 30.0858 &   -27 47 57.053 &24.77  &25.53 & 0.29   &0.11   &0.04\\
s15&  1120.9& 4920.6& 3 32 48.3373 &   -27 47 38.937 &24.95  &24.55 & 0.88   &0.17   &0.04\\
s16&  846.4 & 5071.1& 3 32 48.9577 &   -27 47 34.417 &24.97  &25.34 & 0.92   &0.22   &0.04\\
s17*&  9372.6& 5878.6& 3 32 29.6816 &   -27 47 10.277 &24.32  &23.57 & 1.22   &0.99   &0.31\\
s18&  9350.6& 6065.6& 3 32 29.7313 &   -27 47 04.667 &25.01  &24.94 & 1.23   &0.42   &0.21\\
s19&  5224.5& 7207.5& 3 32 39.0581 &   -27 46 30.390 &23.57  &24.13 & 0.84   &0.31   &0.17\\
s20*&  7461.9& 7446.6& 3 32 34.0005 &   -27 46 23.233 &24.82  &23.95 & 0.54   &0.39   &0.02\\
s21&  4064.1& 8767.1& 3 32 41.6796 &   -27 45 43.589 &24.55  &24.80 & 0.78   &0.59   &0.16\\
s22&  5011.9& 9501.1& 3 32 39.5369 &   -27 45 21.580 &25.36  &24.16 & 0.02   &0.7&0.4\\
s23&  6490.1& 9002.3& 3 32 36.1963 &   -27 45 36.556 &24.60  &22.29 & 0.98   &0.65   &0.29\\
s24&  6662.9& 8585.6& 3 32 35.8060 &   -27 45 49.058 &24.40  &22.93 & 0.30   &0.51   &0.17\\
s25&  6994.6& 8230.3& 3 32 35.0564 &   -27 45 59.719 &24.78  &23.71 & 0.13   &0.66   &0.30\\
s26*&  6431.0& 8852.9& 3 32 36.3300 &   -27 45 41.038 &24.18  &25.10 & 0.16   &0.63   &0.28\\
s27&  5736.6& 9952.6& 3 32 37.8987 &   -27 45 08.041 &23.91  &24.38 & 0.57   &0.85   &0.49\\
s28&  5623.6& 9773.9& 3 32 38.1542 &   -27 45 13.401 &23.93  &22.19 & 1.03   &1.28   &0.48\\
\\
\tableline
\end{tabular}
\tablenotetext{a}{Barred galaxies are indicated by an asterisk}
\tablenotetext{b}{The coordinates of UDF 5898A are $3^{\rm h}
32^{\rm m}31.3973^{\rm s}$, $-27^\circ47^\prime
13.785^{\prime\prime}$; the coordinates of UDF 5898B are $3^{\rm
h}32^{\rm m}31.3967^{\rm s}$, $-27^\circ47^\prime
13.096^{\prime\prime}$}
\end{center}
\end{table}

\newpage
\begin{table}
\begin{center}
\caption{Elliptical Galaxies\label{tab:ell}}
\begin{tabular}{cccccccccccccc}
\tableline\tableline
UDF &&&&&&&&&&&&&\\
\tableline
     8&    25&    36&    68&    97&   100&   153&   176&   191&   206&   221&   287&   310&   582\\
   600&   631&   703&   731&   865&   900&   901&  1088&  1344&  1358&  1453&  1481&  1564&  1607\\
  1727&  1960&  2016&  2063&  2107&  2162&  2201&  2245&  2248&  2293&  2322&  2329&  2387&  2414\\
  2518&  2927&  2974&  3048&  3088&  3136&  3174&  3288&  3332&  3333&  3475&  3591&  3677&  3869\\
  4096&  4142&  4320&  4350&  4389&  4396&  4445&  4481&  4527&  4551&  4587&  4805&  4913&  5187\\
  5263&  5959&  6018&  6022&  6027&  6047&  6288&  6747&  6853&  6853&  7121&  7398&  7780&  8069\\
  8138&  8316&  8680&  9090&  9264&  9532&  9765&  9778&  9847&  9962&     &     &     &     \\
\end{tabular}
\begin{tabular}{lccccccccc}
\tableline\tableline
Number&x&y&RA(2000)&DEC(2000)&$\mu_{\rm I}$&I&B-V&V-i&i-z\\
\tableline
E1  &   4154.8 & 1231.1 & 3 32 41.4815 &   -27 49 28.72 & 24.92 & 26.05 & 0.36&0.62&0.20\\
E2  &   6004.4 & 1086.9 & 3 32 37.2988 &   -27 49 34.02 & 24.75 & 25.42 & 0.40&0.62&-0.19\\
E3  &   9026.7 & 4118.9 & 3 32 30.4637 &   -27 48 03.07 & 24.24 & 23.52 & 1.15&0.49&0.15\\
E4  &   3674.5 & 8546.7 & 3 32 42.5604 &   -27 45 50.20 & 23.56 & 22.36 & 0.28&1.04&0.18\\
E5  &   5741.9 & 9595.4 & 3 32 37.8870 &   -27 45 18.76 & 23.80 & 23.79 & 0.35&0.86&0.29\\
E6  &   4826.6 & 9687.6 & 3 32 39.9555 &   -27 45 15.98 & 25.13 & 25.75 & 0.60&0.30&0.14\\
\tableline
\end{tabular}
\end{center}
\end{table}

\newpage
\begin{table}
\begin{center}
\caption{Lyman Break Drop-out
Galaxies\tablenotemark{a}\label{tab:lbg}}
\begin{tabular}{lclclc}
\tableline\tableline
B-band drop-outs&&&&&\\
\tableline
UDF &SFR\tablenotemark{b}&UDF&SFR\tablenotemark{b}&UDF&SFR\tablenotemark{b}\\
\tableline
    65(C)&7.9&
       141(T)&9.5&
          401(C)&21\\
             631(E)&8.4&
                985(D)&5.7&
                  2394(C)&3.7\\
                    3001(C)&2.0&
                    3458+3418(C)&13&
                      3778(CC)&19\\
                        4313(D)&4.1&
                          4548(T)&7.7&
                            4551(E)&16\\
                              4685(D)&14&
                              4765+4795(CC)&21&
                                5548(CC)&12\\
                                  6137(D)&2.3&
  6209(D)&1.5&
    6450(C)&3.8\\
      6543(D)&15&
        6709(C)&1.1&
          6808(D)&4.8\\
            7036(S)&77&
              7044(T)&3.1&
                8092(C)&16\\
                  8419(D)&6.6&
                   9085+8865(C)&9.7&
                    9310(T)&1.8\\
                    C7(C) &7.1&
                    E19.47(E)&15&
                    CC11(CC)&61\\
\tableline
V-band drop-outs&&&&&\\
\tableline
1796(D)&6.2&
2350(CC)&42&
2881(T)&58\\
3377+3398(D)&30&
5225(C)&48&
5928(D)&6.5\\
6139(D)&63&
6681(D)&24&
7328(CC)&465\\
8326(D)&18&
8664(D)&23&
8682(CC)&50\\
D11(D)&9.9&&&&\\

\end{tabular}
\tablenotetext{a}{initials for each morphological type are given:
C=chain, CC=clump cluster, D=double, T=tadpole, S=spiral, E=elliptical}
\tablenotetext{b}{Star formation rates in M$_\odot$ yr$^{-1}$}
\end{center}
\end{table}

\newpage

\newpage

\newpage

\newpage
\begin{figure}\epsscale{1}\plotone{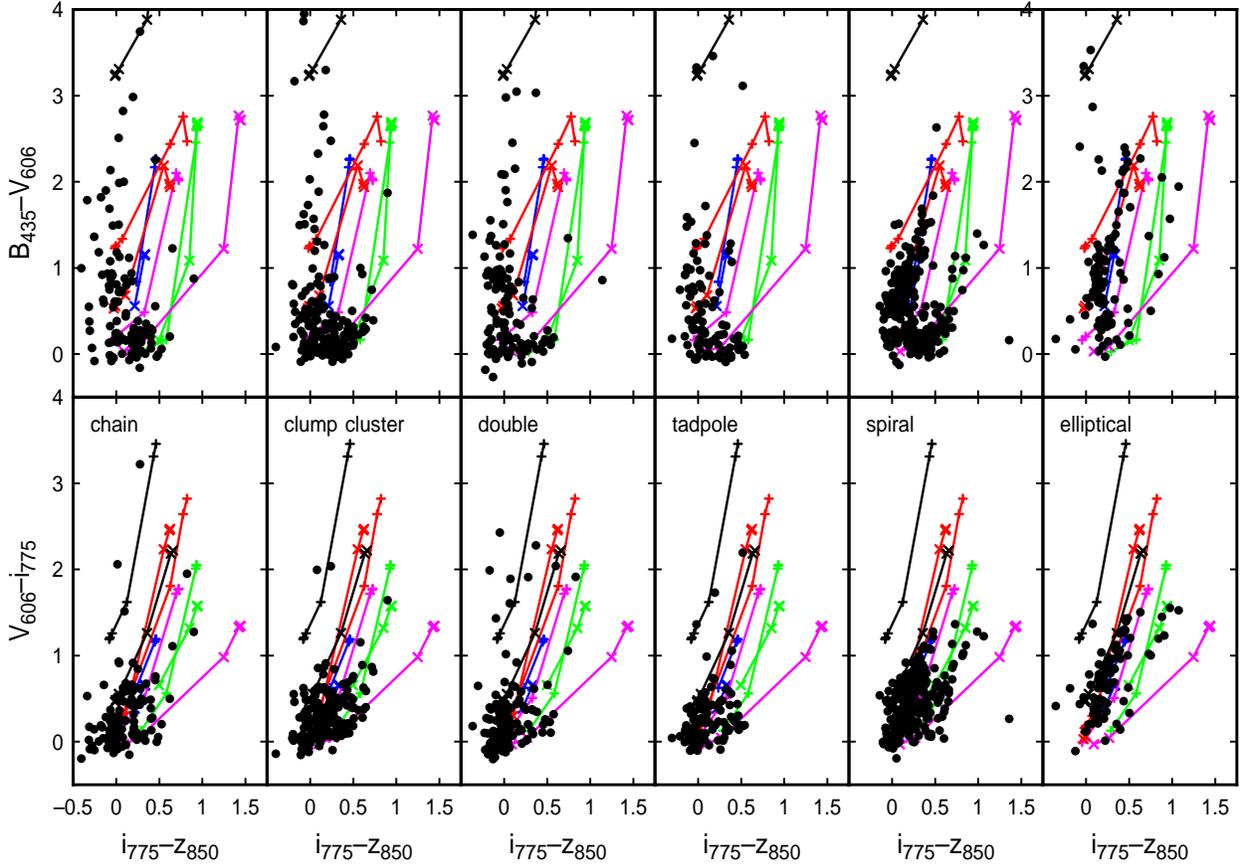}\caption{Color-color
distributions are shown for all 884 galaxies divided by
morphology. Superposed curves are stellar evolution models for
redshifts 0 (blue x symbol), 0.5 (blue $+$), 1 (green x), 1.5
(green $+$), 2 (magenta x), 2.5 (magenta $+$), 3 (red x), 3.5 (red
$+$), 4 (black x), and 4.5 (black $+$). The models assume star
formation began at $z=6$ and decayed exponentially; the curves
trace out the colors as the e-folding time in Gy varies as 0.01,
0.03, 0.1, 0.3, 1, 3, and infinity (high decay times are the most
blue).} \label{fig:overlay}\end{figure}

\newpage
\begin{figure}\epsscale{1}\plotone{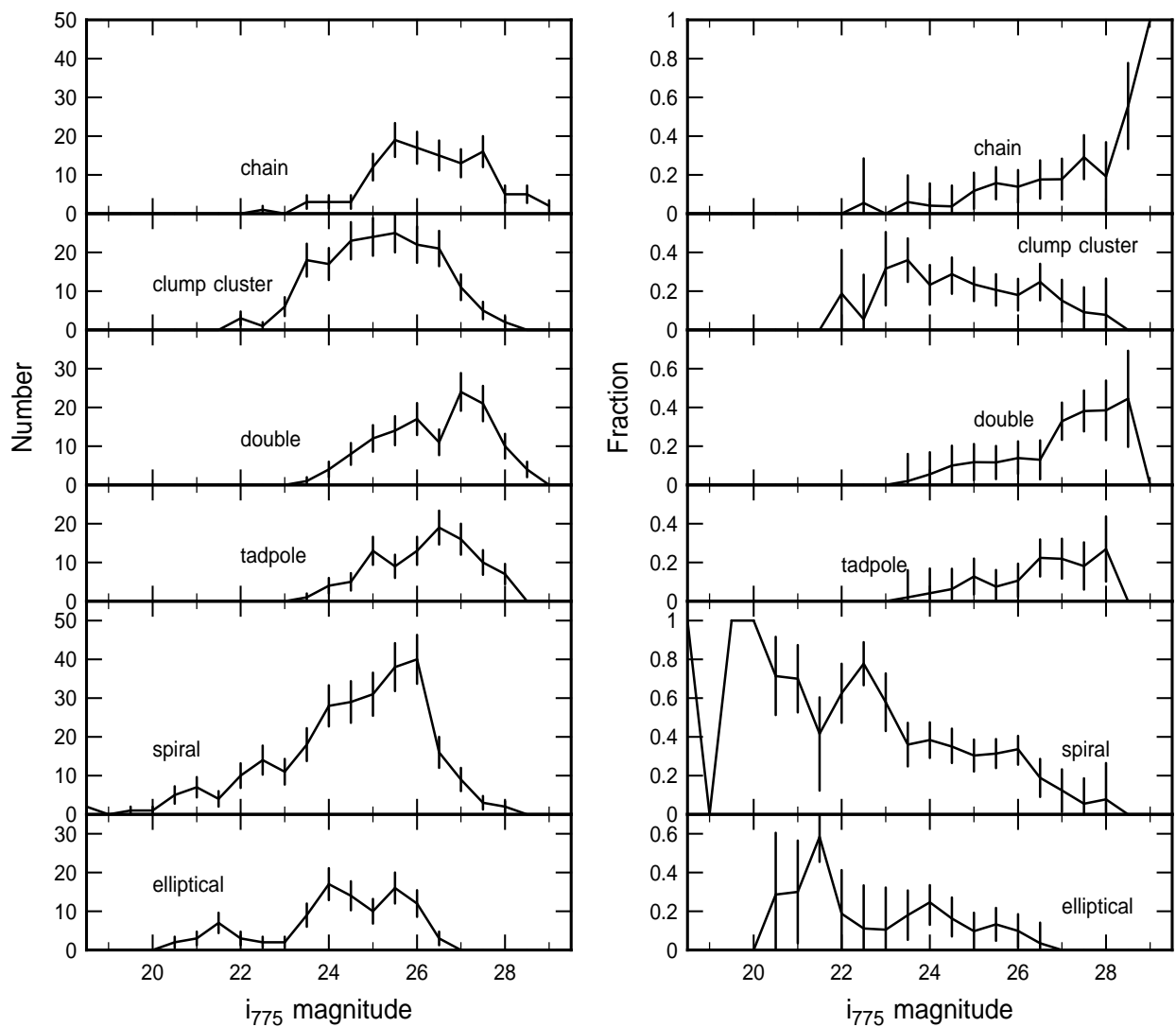}\caption{The
number and fraction of each galaxy type is shown as a function of
apparent magnitude, with $1\sigma$ error bars. Linear structures
dominate at the faintest
magnitudes.}\label{fig:fractions}\end{figure}
\newpage
\begin{figure}\epsscale{0.8}\plotone{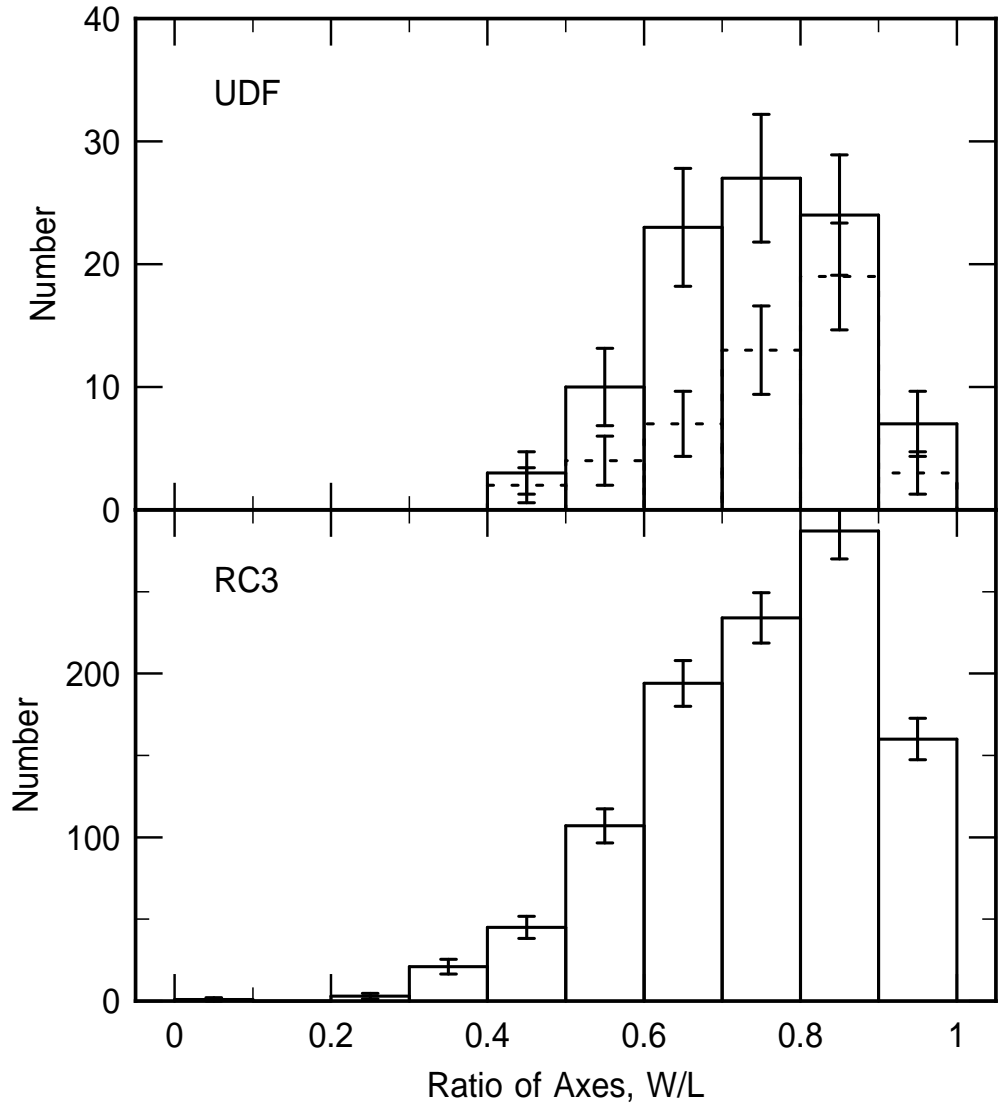}\caption{The distribution of
the ratio of axis for elliptical galaxies in the UDF compared to
the RC3. The dotted histogram for the UDF is for the half with the
highest surface brightness. The UDF distribution resembles the
local distribution, suggesting that elliptical galaxy shapes are
standard and were probably determined relatively quickly in the
early Universe.}\label{fig:ellfractwl}\end{figure}

\newpage
\begin{figure}\epsscale{1}\plotone{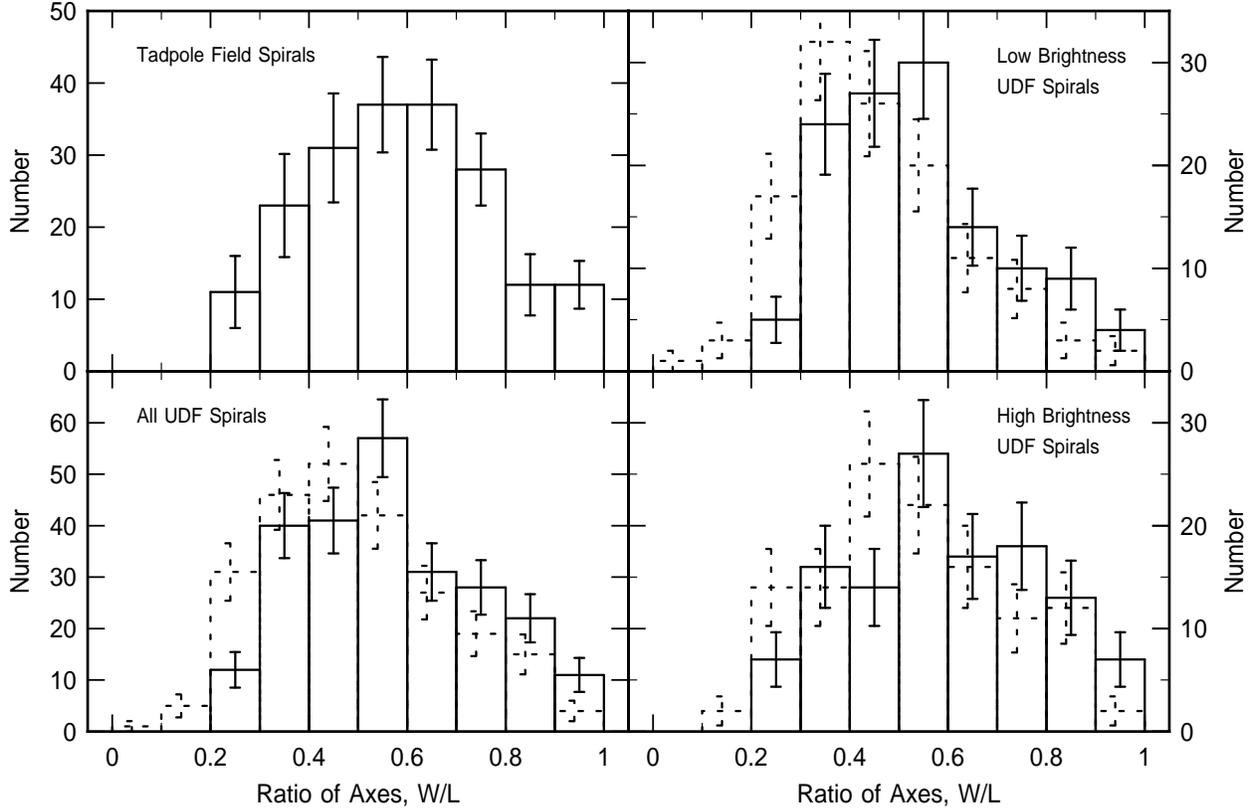}\caption{The distribution of
the ratio of axis (width to length, $W/L$) for spiral galaxies in
the deep field of the Tadpole Galaxy and in the UDF field.  The
distributions differ significantly from the local distributions
(Fig. 9) because of a lack of face-on spirals at high redshift.
Randomly oriented disks would have a flat distribution in a
diagram like this.  The fall-off at low $W/L$ results from the
intrinsic disk thickness. The position of this fall-off is
significantly larger at high redshift than for local galaxies,
indicating that the high redshift galaxies have thicker disks.}
\label{fig:spiralwl}\end{figure}

\newpage
\begin{figure}\epsscale{1}\plotone{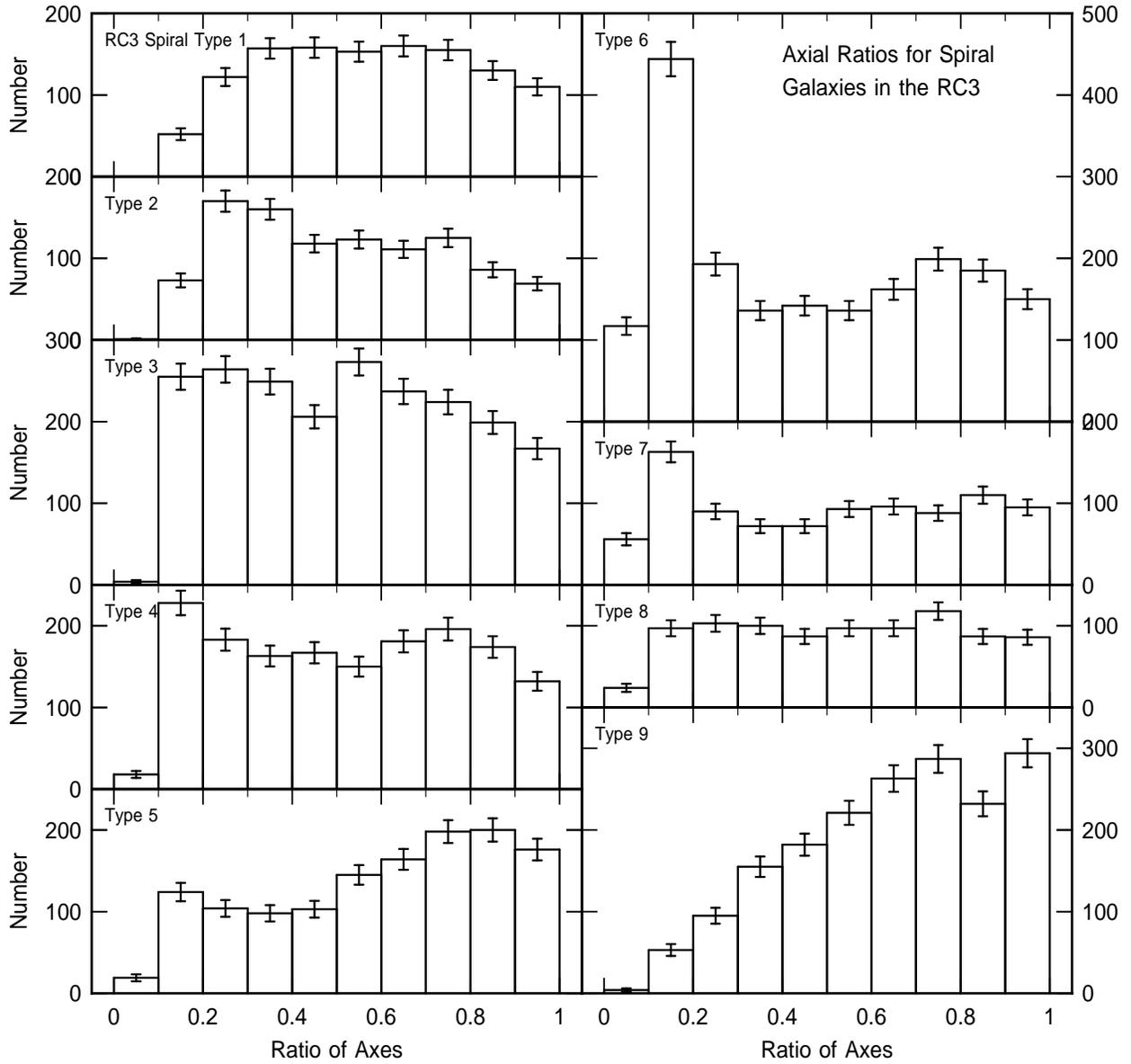}\caption{The distribution of
the ratio of axis for spiral galaxies in the RC3, divided
according to de Vaucouleurs type. }
\label{fig:spiralwlrc3}\end{figure}

\newpage
\begin{figure}\epsscale{1}\plotone{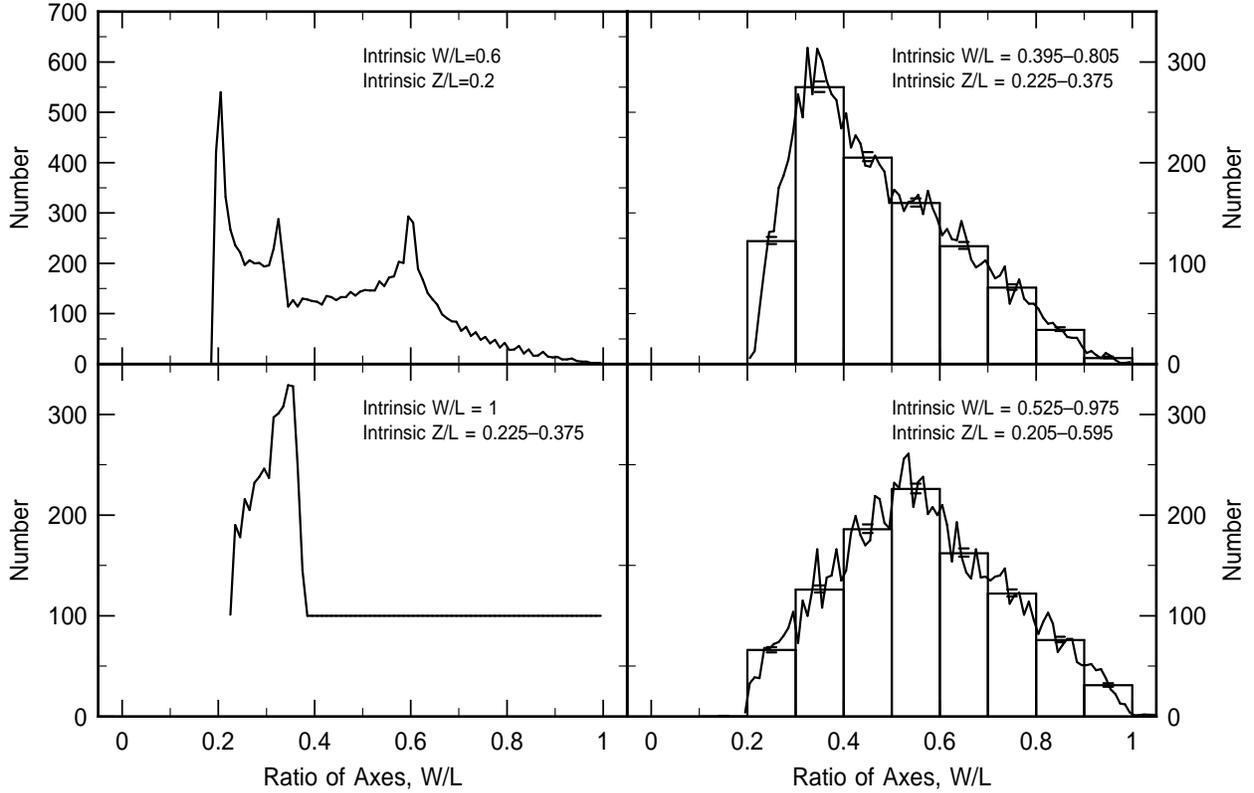}\caption{Models of
distribution of axial ratio for intrinsically triaxial galaxies
viewed at random angles, demonstrating how the observed deviations
from a uniform distribution could arise at high redshift without
surface brightness selection effects. The range of intrinsic width
to length ($W/L$) and thickness to length ($Z/L$) is indicated on
each panel. The two distributions on the right illustrate how the
centrally peaked and skewed distributions of the observed spirals
can be matched by a triaxial model.} \label{fig:wl}\end{figure}

\newpage
\begin{figure}\epsscale{0.8}\plotone{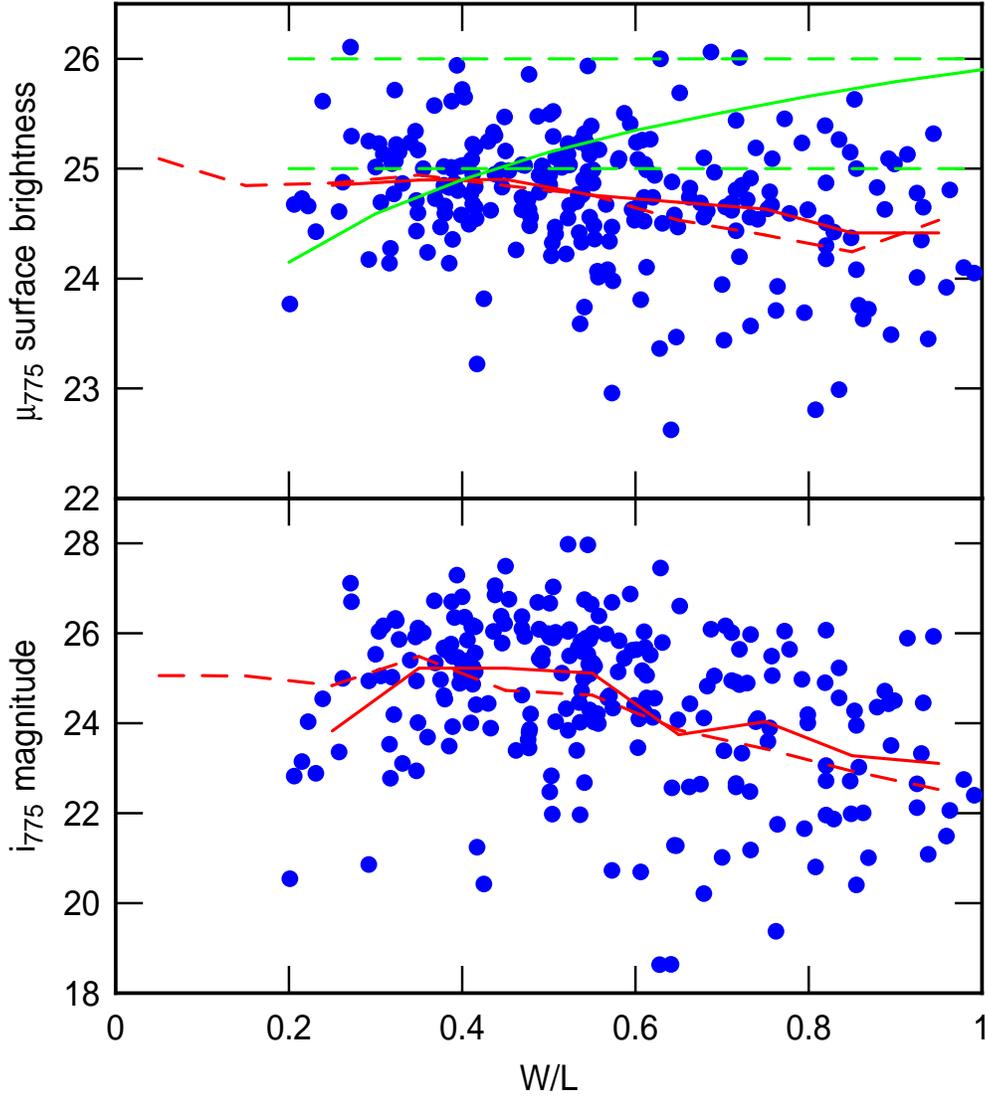}\caption{The
distribution of magnitude and surface brightness for all of the
spiral galaxies in this catalog, plotted versus the apparent ratio
of axes, $W/L$.  The plotted points and the solid red lines use
the axial ratios tabulated in the on-line UDF catalog, while the
dashed red lines use the ratio of axes determined by the authors
from individual ellipse fits on IRAF. In the bottom panel. the
edge-on galaxies are fainter as a result of their lower projected
areas. In the top panel, the dashed green lines represent the
range of surface brightnesses where galaxies begin to drop below
the detection limit. The solid green line is a model where the
surface brightness is proportional to the path length through the
disk.  As the path length decreases for more face-on systems, and
the average surface brightness becomes fainter, the detection
limit is passed and the galaxies disappear from view, making the
density of plotted points lower.} \label{fig:wlvsmag}\end{figure}

\newpage
\begin{figure}\epsscale{1.}\plotone{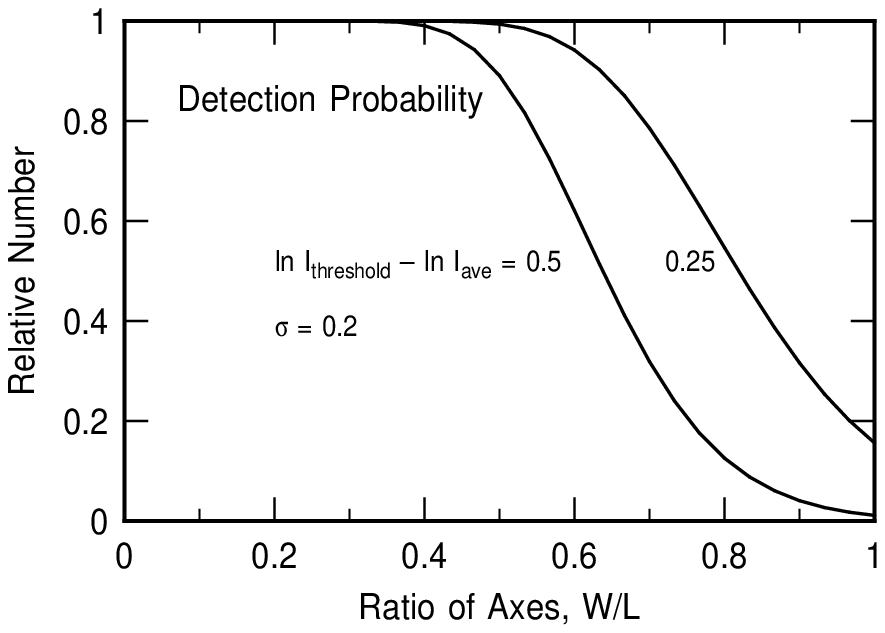}
\caption{A model for the detection probability of spirals which
have a distribution of face-on surface brightnesses that is
Gaussian in magnitude (i.e., log-normal in intensity) with a
dimensionless dispersion $\sigma=0.2$ and a value at the peak,
$I_{\rm ave}$, fainter than the survey threshold, $I_{\rm
threshold}$ by factors of $e^{0.5}$ and $e^{0.25}$. These two
cases bracket the observed fall-off in galaxy counts for the range
of $W/L$ between 0.5 and 1 (as in Fig. 8). }
\label{fig:faceon}\end{figure}


\clearpage

\begin{thebibliography}{}

\bibitem[]{1041} Abraham, R., Tanvir, N., Santiago, B., Ellis, R.,
Glazebrook, K., \& van den Bergh, S. 1996a, MNRAS, 279, L47

\bibitem[]{1044}Abraham, R., van den Bergh, S., Glazebrook, K.,
Ellis, R., Santiago, B., Surma, P., \& Griffiths, R. 1996b, ApJS,
107, 1

\bibitem[]{1048} Adelberger, K., Steidel, C., Shapley, A., Hunt, M.,
Erb, D., Reddy, N., \& Pettini, M. 2004, ApJ, 607, 226

\bibitem[]{1051} Athanassoula, E., \& Bosma, A. 2003, ApSpSci, 284,
491

\bibitem[]{1054}Beckwith, S.V.W. et al., 2005, in preparation

\bibitem[]{} Binggeli, B., \& Popescu, C. C. 1995, A\&A, 298, 63

\bibitem[]{} Bouwens, R.J., Illingworth, G.D., Blakeslee, J.P.,
Broadhurst, T.J., \& Franx, M. 2004 ApJ, 611, L1

\bibitem[]{} Brinchmann, J., et al. 1998, ApJ, 499, 112

\bibitem[]{1056} Bruzual, G. \& Charlot, S. 2003, MNRAS, 344, 1000

\bibitem[]{1058}Bunker, A., Stanway, E., Ellis, R., \& McMahon,
R.2004, MNRAS, 355, 374

\bibitem[]{1061}Conselice, C. 2004, astro-ph/0405102

\bibitem[]{1066} Conselice, C., Grogin, N.A., Jogee, S., Lucas, R.A.,
Dahlen, T., de Mello, D., Gardner, J.P., Mobasher, B.,
Ravindranath, S. 2004, ApJ, 600, L139

\bibitem[]{1070}Cowie, L., Hu, E., \& Songaila, A. 1995, AJ, 110,
1576

\bibitem[]{} Dalcanton, J.J., \& Schectman, S.A. 1996, ApJ, 465,
L9

\bibitem[]{} Dickinson, M. 2000,
Philosophical Transactions of the Royal Society of London, Series
A, Vol. 358, p.2001

\bibitem[]{1073}Elmegreen, B.G., Elmegreen, D.M., \& Hirst, A.C.
2004b, ApJ, 612, 191 (Paper V)

\bibitem[]{1076} Elmegreen, B.G., \& Elmegreen, D.M. 2005, ApJ,
627, 632 (Paper III)

\bibitem[]{1079}Elmegreen, D.M., Elmegreen, B.G., \& Sheets, C.M.
2004, ApJ, 603, 74 (Paper I)

\bibitem[]{1082}Elmegreen, D.M., Elmegreen, B.G., \& Hirst, A.C.
2004a, ApJ, 604, L21 (Paper II)

\bibitem[]{1085}Elmegreen, D.M., Elmegreen, B.G., \& Ferguson, T.E.,
2005, ApJL, 623, L71 (Paper IV)

\bibitem[]{1088} Ferguson, H. C., Dickinson, M., \& Williams, R.
2000, ARA\&A, 38, 667

\bibitem[]{1091} Franceschini, A., Silva, L., Fasano, G., Granato,
G.L., Bressan, A., Arnouts, S., Danese, L. 1998, ApJ, 506, 600

\bibitem[]{1094} Freeman, K.C. 1970, ApJ, 160, 811

\bibitem[]{1096}Giavalisco, M. et al. 2004, ApJ, 600, 103

\bibitem[]{1098}Immeli, A., Samland, M., Westera, P., \& Gerhard, O.
2004, ApJ, 611, 201

\bibitem[]{1101} Jogee, S. et al. 2004, ApJL, 615, 105

\bibitem[]{1103}Madau, P., Pozzetti, L., \& Dickinson M., 1998, ApJ,
106, 116

\bibitem[]{} O'Neil, K., Bothun, G.D., \& Impey, C.D. 2000, ApJS,
128, 99

\bibitem[]{} Reshetnikov, V., Battaner, E., Combes, F., \& Jim\'enez-Vicente, J.
2002, A\&A, 382, 513

\bibitem[]{1106}Reshetnikov, V., Dettmar, R.-J., \& Combes, F. 2003,
A\&A, 399, 879

\bibitem[]{1109}Rhoads, J. et al. 2005, ApJ, 621, 582

\bibitem[]{} Sheth, K., Regan, M.W., Scoville, N.Z., \& Strubbe, L.E. 2003,
ApJ, 592, 13

\bibitem[]{} Somerville, R.S., Primack, J.R., Faber, S.M. 2001,
MNRAS, 320, 504

\bibitem[]{1111} Steidel, C., Adelberger, K., Giavalisco, M.,
Dickinson, M., \& Pettini, M. 1999, ApJ, 519, 1

\bibitem[]{1114} Straughn, A., Ryan, E., Cohen, S., Hathi, N.,
Windhorst, R., \& Pasquali, A. 2004, BAAS, 205, 9417

\bibitem[]{} Sung, E.-C., Han, C., Ryden, B.\ S.,
Patterson, R.\ J., Chun, M.-S., Kim, H.-I., Lee, W.-B., \& Kim,
D.-J. 1998, ApJ, 505, 199

\bibitem[]{} Thompson, R. et al. 2005, astroph/053504

\bibitem[]{1117} Tran, H. et al. 2003, ApJ, 585, 750

\bibitem[]{1119} Tremblay, B. \& Merritt, D. 1996, AJ, 111, 2243

\bibitem[]{1121} Tully, R. B., \& Fisher, J. R. 1977, A\&A, 54, 661

\bibitem[]{} van den Bergh, S. 2002, PASP, 114, 797

\bibitem[]{1123}van den Bergh, S., Abraham, R.G., Ellis, R.S.,
Tanvir, N.R., Santiago, B.X., \& Glazebrook, K.G. 1996, AJ 112,
359

\bibitem[]{} van den Bergh, S., Cohen, J.G., Hogg, D. W., \& Blandford,
R. 2000, AJ, 120, 2190

\bibitem[]{} van den Bergh, S., Abraham, R.G., Whyte,
L.F., Merrifield, M.R., Eskridge, P.B., Frogel, J.A., \& Pogge, R.
2002, AJ, 123, 2913

\bibitem[]{1127} de Vaucouleurs, G., de Vaucouleurs, A., Corwin, H.,
Buta, R., Paturel, G., \& Fouque, P. 1991, Third Reference
Catalogue of Galaxies,  New York: Springer-Verlag

\bibitem[]{1131}Volonteri, M., Saracco, P., \& Chincarini, G. 2000,
A\&AS, 145, 111

\bibitem[]{1134}Williams, R., et al. 1996, AJ, 112, 1335

\bibitem[]{1136} Yan, H., \& Windhorst, R.A. 2004, ApJL, 612, 93

\end{thebibliography}
\end{document}